\documentclass[10pt,english,aps,p9,notitlepage,nofootinbib]{revtex4-1} 
\usepackage[T1]{fontenc}
\usepackage[latin9]{inputenc}
\usepackage{amsmath}
\usepackage{esint}
\usepackage{graphicx}
\usepackage{hyperref} 
\usepackage[caption=false]{subfig}

\def\eq#1{(\ref{#1})}

\makeatletter
\@ifundefined{textcolor}{}
{%
 \definecolor{BLACK}{gray}{0}
 \definecolor{WHITE}{gray}{1}
 \definecolor{RED}{rgb}{1,0,0}
 \definecolor{GREEN}{rgb}{0,1,0}
 \definecolor{BLUE}{rgb}{0,0,1}
 \definecolor{CYAN}{cmyk}{1,0,0,0}
 \definecolor{MAGENTA}{cmyk}{0,1,0,0}
 \definecolor{YELLOW}{cmyk}{0,0,1,0}
}

\usepackage[svgnames,rgb]{xcolor}

\usepackage{ifpdf}
\ifpdf

\IfFileExists{lmodern.sty}
 {\usepackage{lmodern}}{}

\fi % end if pdflatex is used

\makeatother

\usepackage{babel}
\begin{document}
\global\long\def\ave#1{\left\langle #1 \right\rangle }
\global\long\def\absol#1{\left| #1 \right|}
\global\long\def\mev{{\rm \, MeV}}
\global\long\def\gev{{\rm \, GeV}}
\global\long\def\dpp#1{\frac{d^{3}#1}{(2\pi)^{3}}}
\global\long\def\mh{\hat{\mu}}

\title{Cumulants and Correlation Functions vs the QCD phase diagram
}

\author{Adam Bzdak}
\email{bzdak@fis.agh.edu.pl}
\affiliation{AGH University of Science and Technology,\\
Faculty of Physics and Applied Computer Science,\\ 
30-059 Krak\'ow, Poland}  

\author{Volker Koch}
\email{vkoch@lbl.gov}
%\homepage{http://www-nsdth.lbl.gov/~vkoch}
\affiliation{Nuclear Science Division,\\
Lawrence Berkeley National Laboratory,\\
Berkeley, CA, 94720, USA}

\author{Nils Strodthoff}
\email{nstrodthoff@lbl.gov}
\affiliation{Nuclear Science Division,\\
Lawrence Berkeley National Laboratory,\\
Berkeley, CA, 94720, USA}

\begin{abstract}
In this paper we discuss the relation of particle number cumulants
and correlation functions. It is argued that measuring couplings 
of the genuine multi-particle correlation functions could provide cleaner information 
on possible non-trivial dynamics in heavy-ion collisions. We extract integrated
multi-proton correlation functions from the presently available experimental data on proton cumulants. 
We find that the STAR data contain significant four-proton correlations,
at least at the lower energies, with indication of changing dynamics 
in central collisions. We also find that these correlations
are rather long-ranged in rapidity.
Finally, using the Ising model, we demonstrate
how the signs of the multi-proton correlation functions may be used
to exclude certain regions of the phase diagram close to the critical point.
 
\end{abstract}
\maketitle

\section{Introduction}

The search for structures in the QCD phase diagram, such as a critical point
or a first order phase coexistence region has been at the forefront of
strong interaction research for the last several years. Most experimental
and theoretical effort in this regard has concentrated on the measurement
and calculation of cumulants of conserved charges, in particular of baryon
number cumulants \cite{Stephanov:1998dy,Stephanov:2008qz,Skokov:2010uh,Friman:2011pf,
Kitazawa:2013bta,Adamczyk:2013dal,Adamczyk:2014fia},
see, e.g., \cite{Koch:2008ia} for an overview. Different ideas, based on an intermittency analysis in the
transverse momentum phase space are also explored \cite{Anticic:2009pe,Anticic:2012xb}.

Cumulants of the particle number distribution have the advantage that they
are easily accessible in finite temperature field theory since they are
simply given by derivatives of the free energy with respect to an
appropriate chemical potential. However they have the disadvantage that they
mix correlations of different order. For example in case of a system of
uncorrelated particles of one species, say protons, governed by the
Poisson distribution, all cumulants are given by the mean number of
particles, $K_{i}=\langle N\rangle$ for all $i$. Similarly, for
system of uncorrelated resonances which decay in two particles, the
cumulants are simply given by $K_{i}=2^{i}\langle N_\text{res}\rangle $, 
with $\langle N_\text{res}\rangle $ the average number of
resonances. Therefore, a large value for the forth order cumulant does not
necessarily mean the presence of strong four-particle correlations (in our illustrative case we have only two-particle correlations).
Consequently, the fact that STAR sees a cumulant ratio for protons of $%
K_{4}/K_{2}\simeq 3.5$ at $\sqrt{s}=7.7\gev$ \cite{Luo:2015ewa} may well be the result of strong two particle
correlations, rather than three and four body correlations, which would be
expected close to a critical point \cite{Stephanov:2008qz}.

Therefore, it would be very valuable if the true correlation functions could
be extracted from the measured cumulants. In this paper we will discuss
how this can be done, at least for the case of one species of particles,
such as protons (see also \cite{Ling:2015yau}). For net-proton cumulants, i.e.\ cumulants of the difference
distribution of protons and anti-protons, this is unfortunately not the
case. However, at the beam energies where STAR sees the strongest deviation
from Poisson behavior, the number of anti-proton to proton ratio is
vanishingly small and thus the anti-protons can be ignored.

This paper is organized as follows. First we demonstrate how the true
correlation functions can be related to the cumulants, then we apply these
relations to the preliminary STAR data. Next we discuss the 
centrality, rapidity, and energy dependence of these correlation functions. Finally we illustrate
how just the information about the signs of these correlation functions can
be used to exclude certain regions around the critical point.

\section{Cumulants and Correlations Functions}

Let us start by introducing the correlation functions, beginning with two
particles. The two particle density for particles with momenta $p_{1}$ and $%
p_{2}$, $\rho _{2}\left( p_{1},p_{2}\right) $, is given by%
\begin{equation}
\rho _{2}(p_{1},p_{2})=\rho _{1}(p_{1})\rho _{1}(p_{2})+C_{2}(p_{1},p_{2}),
\label{eq:rho-2}
\end{equation}%
where $\rho _{1}\left( p\right) $ refers to the one particle density, and $%
C_{2}(p_{1},p_{2})$ represents the two-particle correlation function. 

In general the two
particle density and correlation function depend on the momenta of both
particles. In the following, we will restrict ourselves to correlations in
rapidity and adopt the following notation 
\begin{eqnarray}
\rho _{2}\left( y_{1},y_{2}\right) &=&\int dp_{t,1}d\phi _{1}dp_{t,2}d\phi
_{2}\rho _{2}\left( p_{1},p_{2}\right) ,  \notag \\
C_{2}\left( y_{1},y_{2}\right) &=&\int dp_{t,1}d\phi _{1}dp_{t,2}d\phi
_{2}C_{2}\left( p_{1},p_{2}\right) ,  \notag \\
C_{2} &=&\int dy_{1}dy_{2}C_{2}\left( y_{1},y_{2}\right) ,
\end{eqnarray}%
and similarly for higher order particle densities and correlation functions.

Integrating $\rho _{2}(p_{1},p_{2})$ over the momenta we obtain 
\begin{equation}
F_{2}\equiv \left\langle N\left( N-1\right) \right\rangle =\int
dp_{1}dp_{2}\,\rho _{2}(p_{1},p_{2})=\left\langle N\right\rangle ^{2} + C_{2},
\label{eq:F2}
\end{equation}%
where $N$ is the number of particles under consideration and $C_2$ is the integrated two-particle correlation function. In the absence of correlations, $C_{2}(p_{1},p_{2})=0$, 
we obtain $\langle N^{2}\rangle -\langle N\rangle ^{2}=\langle N\rangle $.

The three particle density depends on the single-particle densities as
well as the two and three-particle correlation functions 
\begin{eqnarray}
\rho _{3}(y_{1},y_{2},y_{3}) &=&\rho _{1}(y_{1})\rho _{1}(y_{2})\rho
_{1}(y_{3})+\rho _{1}(y_{1})C_{2}(y_{2},y_{3})+\rho
_{1}(y_{2})C_{2}(y_{1},y_{3})  \notag \\
&&+\,\rho _{1}(y_{3})C_{2}(y_{1},y_{2})+C_{3}(y_{1},y_{2},y_{3}),
\label{rho-3}
\end{eqnarray}%
and is related to the third order factorial moment $F_{3}=\left\langle
N\left( N-1\right) \left( N-2\right) \right\rangle $ via%
\begin{equation}
F_{3}=\int dy_{1}dy_{2}dy_{3}\rho _{3}\left( y_{1},y_{2},y_{3}\right)
=F_{1}^{3}+3F_{1}C_{2}+C_{3},  
\label{eq:F3}
\end{equation}%
where $C_{3}$ is the integrated genuine three-particle correlation 
function\footnote{The correlation functions $C_{n}$ are often referred to as ``factorial cumulants'' \cite{Ling:2015yau}.} 
and $F_1=\langle N\rangle$.
Similarly the higher order factorial moments are given by\footnote{%
See, e.g., Ref. \cite{Bzdak:2015dja} for explicit definitions of higher order correlation
functions.}%
\begin{eqnarray}
F_{4} &=&F_{1}^{4}+6F_{1}^{2}C_{2}+4F_{1}C_{3}+3C_{2}^{2}+C_{4},
\label{eq:F4} \\
F_{5}
&=&F_{1}^{5}+5F_{1}C_{4}+10F_{1}^{2}C_{3}+10F_{1}^{3}C_{2}+15F_{1}C_{2}^{2}+10C_{2}C_{3}+C_{5},
\label{eq:F5}
\\
F_{6}
&=&F_{1}^{6}+6F_{1}C_{5}+15F_{1}^{2}C_{4}+20F_{1}^{3}C_{3}+15F_{1}^{4}C_{2}+60F_{1}C_{2}C_{3}+45F_{1}^{2}C_{2}^{2}+15C_{2}C_{4}+10C_{3}^{2}+15C_{2}^{3}+C_{6}.
\label{eq:F6}
\end{eqnarray}

Before we discuss the connection between the integrated
  correlation functions and the cumulants, for completeness let us
  discuss a more formal way of calculating multi-particle integrated
  correlation functions. The above formulas connect the factorial
  moments $F_i$ with the integrated correlation functions $C_n$. For
  example, $C_2 = F_2 - F_{1}^2$ (see Eq. (\ref{eq:F2})), which is
  simply $\langle N(N-1)\rangle - \langle N\rangle^2$. In other words,
  the integrated correlation functions can be expressed in terms of the factorial moments of the multiplicity distribution. Suppose that the particles under consideration are characterized by the multiplicity distribution $P(N)$, where $N$ in our case is the number of protons.
The factorial moment $F_k=\langle N!/(N-k)!\rangle$ is conveniently calculated using the generating function $H(z)$
\begin{equation}
F_{k}=\left. \frac{d^{k}}{dz^{k}}H(z)\right| _{z=1},\qquad
H(z)=\sum\nolimits_{N}P(N)z^{N}, \qquad H(1)=1,
\end{equation}
and the integrated correlation function is given by analogous derivatives from the logarithm of $H(z)$.
\begin{equation}
C_{n}=\left. \frac{d^{n}}{dz^{n}}\ln \left[ H(z)\right] \right| _{z=1} .
\end{equation}
For example, $C_2 = H^{''}(1) - (H^{'}(1))^2 = F_2 - F_{1}^{2}$, in
agreement with Eq.~(\ref{eq:F2}), and it is
straightforward to verify Eqs.~(\ref{eq:F3}-\ref{eq:F6}) as well.

The particle number cumulants, $K_{n}$, can be expressed in
terms of the factorial moments \cite{Bzdak:2012ab},%
\begin{eqnarray}
K_{1} &\equiv &\langle N\rangle =F_{1},  \notag \\
K_{2} &\equiv &\langle (\delta N)^{2}\rangle =F_{1}-F_{1}^{2}+F_{2},  \notag
\\
K_{3} &\equiv &\langle \left( \delta N\right) ^{3}\rangle
=F_{1}+2F_{1}^{3}+3F_{2}+F_{3}-3F_{1}(F_{1}+F_{2}),
\label{eq:K1}
\end{eqnarray}
and 
\begin{eqnarray}
K_{4} &\equiv &\langle \left( \delta N\right) ^{4}\rangle -3\langle (\delta
N)^{2}\rangle ^{2}  \notag \\
&=&F_{1}-6F_{1}^{4}+7F_{2}+6F_{3}+F_{4}+12F_{1}^{2}(F_{1}+F_{2})-3(F_{1}+F_{2})^{2}-4F_{1}(F_{1}+3F_{2}+F_{3}),
\end{eqnarray}%
where $\delta N=N-\langle N\rangle $. Formulas for the higher order
cumulants can be found in Ref. \cite{Bzdak:2012ab}.

We note that in the present paper we are interested in the
  multi-proton correlation functions and thus we consider cumulants
  and correlations 
  for protons only. In the above equations $N$ denotes the number of
  protons and not the net-proton number.

Now we can relate the cumulants in terms of the correlation functions and
the mean particle number $\langle N\rangle = F_{1}$ 
\begin{eqnarray}
K_{2} &=&\ave{N}+C_{2},  \label{eq:K2C} \\
K_{3} &=&\ave{N}+3C_{2}+C_{3},  \label{eq:K3C} \\
K_{4} &=&\ave{N}+7C_{2}+6C_{3}+C_{4},  \label{eq:K4C}
\end{eqnarray}%
and vice versa 
\begin{eqnarray}
C_{2} &=&-\ave{N}+K_{2},  \label{eq:C2} \\
C_{3} &=&2\ave{N}-3K_{2}+K_{3},  \label{eq:C3} \\
C_{4} &=&-6\ave{N}+11K_{2}-6K_{3}+K_{4} .  \label{eq:C4}
\end{eqnarray}

Before we apply the above equations to extract the correlation strength from
the STAR data, let us make a few more remarks concerning these correlation
functions.

It should be clear from Eqs. \eq{eq:C2}-\eq{eq:C4} that as we approach the critical
point, $C_n$ is dominated by $K_n$ which scales with the highest power of the correlation length $\xi$ \cite{Stephanov:2008qz}. Thus, following \cite{Stephanov:2008qz}, $C_{2}\sim \xi^{2}$, $C_{3}\sim \xi^{4.5}$, and $C_{4}\sim \xi^{7}$ close to the critical point.

Frequently in the literature, see, e.g., Ref. \cite{Lisa:2005dd}, one refers to correlation function where the
trivial dependence on the particle density/multiplicity is removed 
\begin{equation}
c_{n}\left( y_{1},...,y_{n}\right) =\frac{C_{n}\left( y_{1},...,y_{n}\right) 
}{\rho _{1}\left( y_{1}\right) \cdots \rho _{1}\left( y_{n}\right) },
\label{eq:coupling_0}
\end{equation}%
which we shall refer to as reduced correlation functions or simply
couplings. For example in terms of the reduced correlation functions the two
particle density would be given as 
\begin{equation}
\rho _{2}\left( y_{1},y_{2}\right) =\rho _{1}\left( y_{1}\right) \rho
_{1}\left( y_{2}\right) \left[ 1+c_{2}\left( y_{1},y_{2}\right) \right] .
\end{equation}%
The reduced correlation functions will prove helpful when studying for
instance the centrality dependence of the correlations.
Integrating Eq. (\ref{eq:coupling_0}) over rapidity we obtain
\begin{equation}
C_{k}=\left\langle N\right\rangle ^{k}c_{k} ,
\label{eq:Cc}
\end{equation}%
where $\ave{N}=\int_{\Delta y} \rho_{1}(y)dy$ depends on the rapidity interval $\Delta y$ and we denote%
\begin{equation}
c_{k}=\frac{\int \rho _{1}\left( y_{1}\right) \cdots \rho _{1}\left(
y_{k}\right) c_{k}\left( y_{1},...,y_{k}\right) dy_{1}\cdots dy_{k}}{\int
\rho _{1}\left( y_{1}\right) \cdots \rho _{1}\left( y_{k}\right)
dy_{1}\cdots dy_{k}}.
\label{eq:coupling}
\end{equation}

Using above definition we can write%
\begin{eqnarray}
K_{2} &=&\left\langle N\right\rangle +\left\langle N\right\rangle ^{2}c_{2} , \label{eq:K2-c2}
\\
K_{3} &=&\left\langle N\right\rangle +3\left\langle N\right\rangle
^{2}c_{2}+\left\langle N\right\rangle ^{3}c_{3} ,\\
K_{4} &=&\left\langle N\right\rangle +7\left\langle N\right\rangle
^{2}c_{2}+6\left\langle N\right\rangle ^{3}c_{3}+\left\langle N\right\rangle
^{4}c_{4} . \label{eq:K4-c234}
\end{eqnarray}%

Finally we should point out that a direct relation between correlation
functions and cumulants can not be established if one considers for example
net-proton cumulants. In this case the additional knowledge of various
factorial moments is required. The relevant formulas are given in the
Appendix.

\subsection{Comments}
\label{sec:comments}

Before we analyze the existing data several comments are warranted.

\begin{itemize}

\item[(i)]  First it would be interesting to see how the correlation functions $C_{n}$ and couplings $c_{n}$ scale with
multiplicity if the correlations originate from independent sources
of correlations, e.g., from resonances/clusters or when A+A is a simple superposition of elementary p+p interactions. This will be useful when studying the centrality dependence of the correlations.

Suppose we have $N_{s}$ sources of particles, each
characterized by the multiplicity distribution $P(n_{i})$. The final
multiplicity distribution is given by
\begin{equation}
P(N)=\sum_{n_{1},n_{2},...,n_{N_{s}}}P(n_{1})P(n_{2})\cdots
P(n_{N_{s}})\delta _{n_{1}+...+n_{N_{s}}-N} .
\end{equation}%
Calculating the factorial moment generating function we obtain%
\begin{equation}
H(z)=\sum\nolimits_{N}P(N)z^{N}=\left(
\sum\nolimits_{n_{1}}P(n_{1})z^{n_{1}}\right) ^{N_{s}}=H_{1}(z)^{N_{s}},
\end{equation}%
where $H_{1}(z)$ is the factorial moment generating function from a single
source. The correlation function, $C_{k}$, is given by%
\begin{equation}
C_{k}=\left. \frac{d^{k}}{dz^{k}}\ln \left[ H(z)\right] \right| _{z=1}=N_{%
\text{s}}\left. \frac{d^{k}}{dz^{k}}\ln \left[ H_{1}(z)\right] \right|
_{z=1}=N_{s}C_{k}^{(\text{source})},
\label{eq:Ck-source}
\end{equation}%
where $C_{k}^{(\text{source})}$ is the correlation from a single
source.\footnote{We note that when the sources are distributed according
  to a Poisson distribution then $H(z)=\exp(\langle N_{s}\rangle
  [H_{1}(z) - 1])$ and consequently $C_k = \langle N_{s}\rangle
  F_{k}^{(\text{source})}$, where $\langle N_{s}\rangle$ is the
  average number of sources and $F_{k}^{(\text{source})}$ is the
  factorial moment of a single source. The scaling of the couplings
  $c_{k}$ given by
  Eq.~(\ref{eq:indep_source}) remains the same.} 
As seen from the above equation $C_k$ scales simply with the number of sources since $C_{k}^{(\text{source})}$, being a property of a single independent source, does not depend on $N_{s}$. Comparing Eq. (\ref{eq:Ck-source}) with Eq. (\ref{eq:Cc}) we obtain
\begin{equation}
c_{k}=\frac{N_{s}}{\left\langle N\right\rangle^{k}}C_{k}^{(\text{source})},
\label{eq:c-Ns-N}
\end{equation}
and assuming that the number of produced protons, $N$, is proportional to the
number of sources we have for the couplings 
\begin{equation}
c_{k}\sim \frac{1}{\left\langle N\right\rangle^{k-1}}.
\label{eq:indep_source}
\end{equation}

This result is rather straightforward. The correlation strength, $c_{k}$, for the whole system
gets diluted once there are many independent sources of
correlations. Suppose we have $N_{s}$ sources which correlate two particles each. Then we have $N=2 N_{s}$ particles and $N_{s}=N/2$ correlated pairs. The total number of pairs is $N(N-1)/2 \simeq 2 N_{s}^{2}$ and thus the number of correlated over all pairs scales like $1/N$. Similarly for triplets (now each source correlates three particles) one gets $N/3$ correlated out of  $N(N-1)(N-2)/3! \simeq N^{3}/3!$ total triplets, leading to $1/N^2$.
The scaling, Eq.~\eqref{eq:indep_source}, is expected, e.g., for resonances / clusters of
particles or when A+A can be decomposed into elementary p+p collisions.

We note that the scaling given by Eq. (\ref{eq:indep_source}) is expected to break down when the sources are not independent. For example when we have one coherent source of correlations, the number of correlated pairs might be proportional to the total number of pairs and $c_k$ could become constant 
\begin{equation}
c_{k}\sim \text{const}.
\label{eq:collective}
\end{equation}
as a function of $N$. It would be definitely interesting to observe such transition (from $1/N^{k-1}$ to const.) in experimental data. In the next section we will argue that this is the case for central
collisions in the preliminary STAR data at the lowest energies.

\item[(ii)] Suppose that indeed $c_{2,3,4}$ are constant or depend only very weakly on the number of produced
protons. In this case the correlations, $C_k=\langle N\rangle^k c_{k}$, increase with the number of particles. One scenario would be that the sources of correlation are correlated themselves or that the sources correlate increasing number of particles, e.g., with increasing $N$ clusters get larger (more particles per cluster) leading to $C_{k}^{(\text{source})}$ depending on $N$, see Eq.~(\ref{eq:c-Ns-N}).
Given only the integrated reduced correlation function, it is impossible to distinguish between these various scenarios. In any case, centrality independence of the couplings, indicate that the increasing number of particles are correlated and, for the lack of a better term, we will refer to this behavior as ``collective''.
In this case the
cumulants, $K_{n}$, explicitly depend on $\langle N\rangle ^{i}$, $i=1,2,...,n$, 
see Eqs. (\ref{eq:K2-c2}-\ref{eq:K4-c234}).
Consequently the cumulant ratios depend on multiplicity which makes
the interpretation of the data 
rather complicated.
For example by changing centrality or energy we obviously change 
$\langle N\rangle $ which may result in nontrivial behavior. For
example if $\langle N\rangle \ll 1$, as in the case of anti-protons
at low energy, the cumulants are dominated by the leading term and the
cumulant ratios are close to $1$ even if couplings carry actually some
nontrivial information. 

We conclude that the cumulant ratios are rather tricky to
interpret if the couplings, $c_k$, are constant as a function of produced protons. It seems that studying correlation functions is more appropriate in this case.

\item[(iii)] Similarly we can make some general observations about the rapidity dependence of
cumulants and their ratios. To this end let us consider two limits and let us 
assume the rapidity density is constant, $dN/dy=\rho_{1}(y)=\text{const}$ in
rapidity window of interest: 

(a) The correlations are local in rapidity and depend only on the relative distances,
$c_{k}(y_{1},\ldots,y_{k}) =c_{k}^{0} \delta(y_{1}-y_{2}) \cdots
\delta(y_{k-1}-y_{k})$. In this case the couplings or reduced
correlation functions,  Eq. (\ref{eq:coupling}), are given by
$c_{n}=c_{n}^{0}/(\Delta y)^{n-1}$, where $\Delta y$ is the range in rapidity under consideration (namely, particles are measured in $-\Delta y/2 < y < \Delta y/2$). Consequently, the correlation functions $C_{n}$ and the cumulants scale linearly with $\Delta y$
\begin{equation}
C_{n} \sim \Delta y \quad \rightarrow \quad  K_{n}\sim \Delta y .
\end{equation}
In this case the cumulant ratios, e.g., $K_{4}/K_{2}$, do not depend on
$\Delta y$.

(b) The other extreme are long-range correlations, where the
correlation functions are constant over the rapidity region of
interest\footnote{In the STAR experiment $|y|<0.5$, which is not
                  particularly long-range in rapidity. Thus a constant 
                  $c_{k}(y_{1},\ldots,y_{k})$ may not be such a strong requirement.}. 
In this case $c_{k}(y_{1},\ldots,y_{k}) =c_{k}^{0}$ and $c_k$, defined in Eq. (\ref{eq:coupling}), 
equals $c_{k}^{0}$. Thus the correlation functions, $C_{n}=\langle N\rangle ^{n}c_{n}$, scale 
with the n-th power of the rapidity interval $\Delta y$
\begin{equation}
C_{n} \sim (\Delta y)^{n} ,
\label{eq:Cn-Delta-y}
\end{equation} 
since $\langle N\rangle \sim \Delta y$. The scaling of the cumulants $K_{n}$ in this case is more subtle since the cumulants depend on correlation functions of various order. For example the fourth order cumulant
\begin{equation}
K_{4}=\left\langle N\right\rangle +7\left\langle N\right\rangle
^{2}c_{2}^{0}+6\left\langle N\right\rangle ^{3}c_{3}^{0}+\left\langle
N\right\rangle ^{4}c_{4}^{0},\qquad \left\langle N\right\rangle
=\left\langle N_{\Delta y=1}\right\rangle \Delta y ,
\label{eq:K4-rapi}
\end{equation}
depends on the correlation functions $C_{2}$ to $C_{4}$ and the dependence of $\Delta y$ is thus a polynomial of up to fourth order in $\Delta y$. Here $\langle N_{\Delta y=1} \rangle$ is the average number of particles in $\Delta y=1$.
In this case the cumulant ratios do depend (in general) on the size of the rapidity window $\Delta y$. We will discuss this issue in more detail in the next section.

Of course things get more complicated if the rapidity density $dN/dy$ is not constant and if the correlation length in rapidity is finite but shorter than $\Delta y$.

\item[(iv)] It is clear that at very low energy the majority of observed
protons originate from the incoming nuclei, and are decelerated to mid rapidity. In the simplest model we may  assume that protons
stop in a given rapidity $\Delta y$ bin with some probability $p$ leading to binomial
distribution
%\footnote{This is equivalent to the global baryon conservation \cite{Bzdak:2012an}.}%
\begin{equation}
P(N)=\frac{B!}{N!(B-N)!}p^{N}(1-p)^{B-N},
\label{eq:stop}
\end{equation}%
where $B$ is the {\em total} number of protons (that potentially can stop in $\Delta y$) and $p B$ is the mean number of protons observed in a given acceptance. We note that the above formula, representing the simplest stopping model, is equivalent to the problem of global baryon conservation \cite{Bzdak:2012an}, when the contribution from anti-protons may be neglected (low energies).
The factorial moment generating function is%
\begin{equation}
H(z)=\sum\nolimits_{N}P(N)z^{N}=\left[ 1-p(1-z)\right] ^{B} ,
\end{equation}%
and the couplings are%
\begin{equation}
c_{2}=-\frac{1}{B},\quad c_{3}=\frac{2}{B^{2}},\quad c_{4}=-\frac{6}{B^{3}}.
\end{equation}

We note that $B$ is changing with centrality and this scenario falls
into the class of independent sources of correlations, since protons stop independently in $\Delta y$.

\item[(v)] It would be interesting to measure correlations and couplings between protons and
anti-protons and how they change with energy and centrality. In the Appendix
we derive suitable formulas, which require the knowledge of additional factorial moments.

\item[(vi)] The preliminary STAR data \cite{Luo:2015ewa} show a comparatively large ratio of the fourth-order over 
second-order cumulant, $K_{4}/K_{2} \simeq 3.5$. Given Eqs.~\eqref{eq:K2C} and \eqref{eq:K4C} this does not imply a priori the presence of any four particle correlations, since for sufficiently large two-particle correlations, $C_{2}\gg \ave{N}$, the cumulant ratio may be as large as $K_{4}/K_{2}\simeq 7$ without any three- and four-particle correlations. 

% This is not very importnt. We need a simple argument and maybe we can add this later.
%(v) If efficiency is given by binomial than couplings are independent on efficiency.

\end{itemize}

\section{Extracting correlation functions from data}

\begin{figure}[t]
\begin{center}
\includegraphics[scale=0.3]{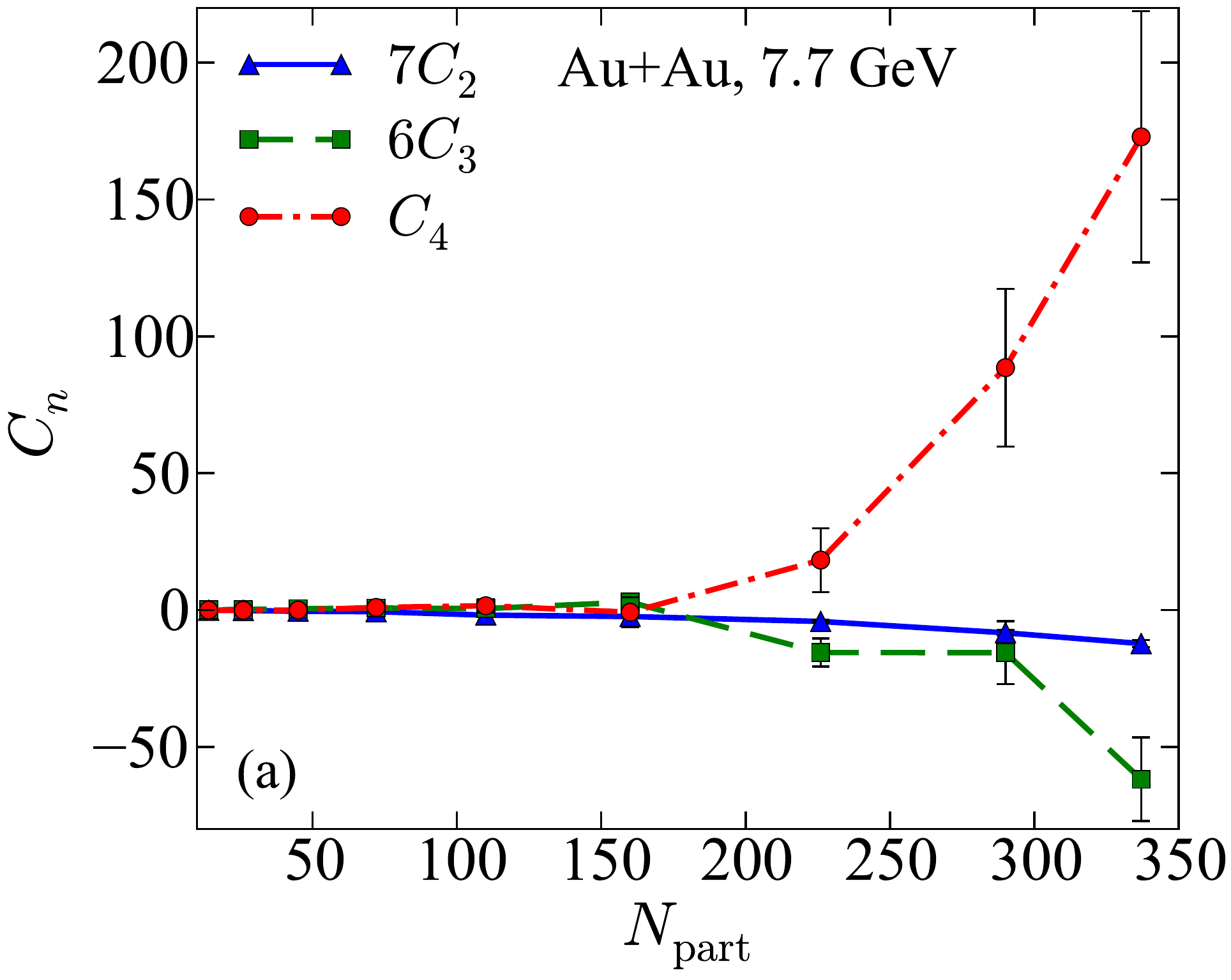}
\includegraphics[scale=0.3]{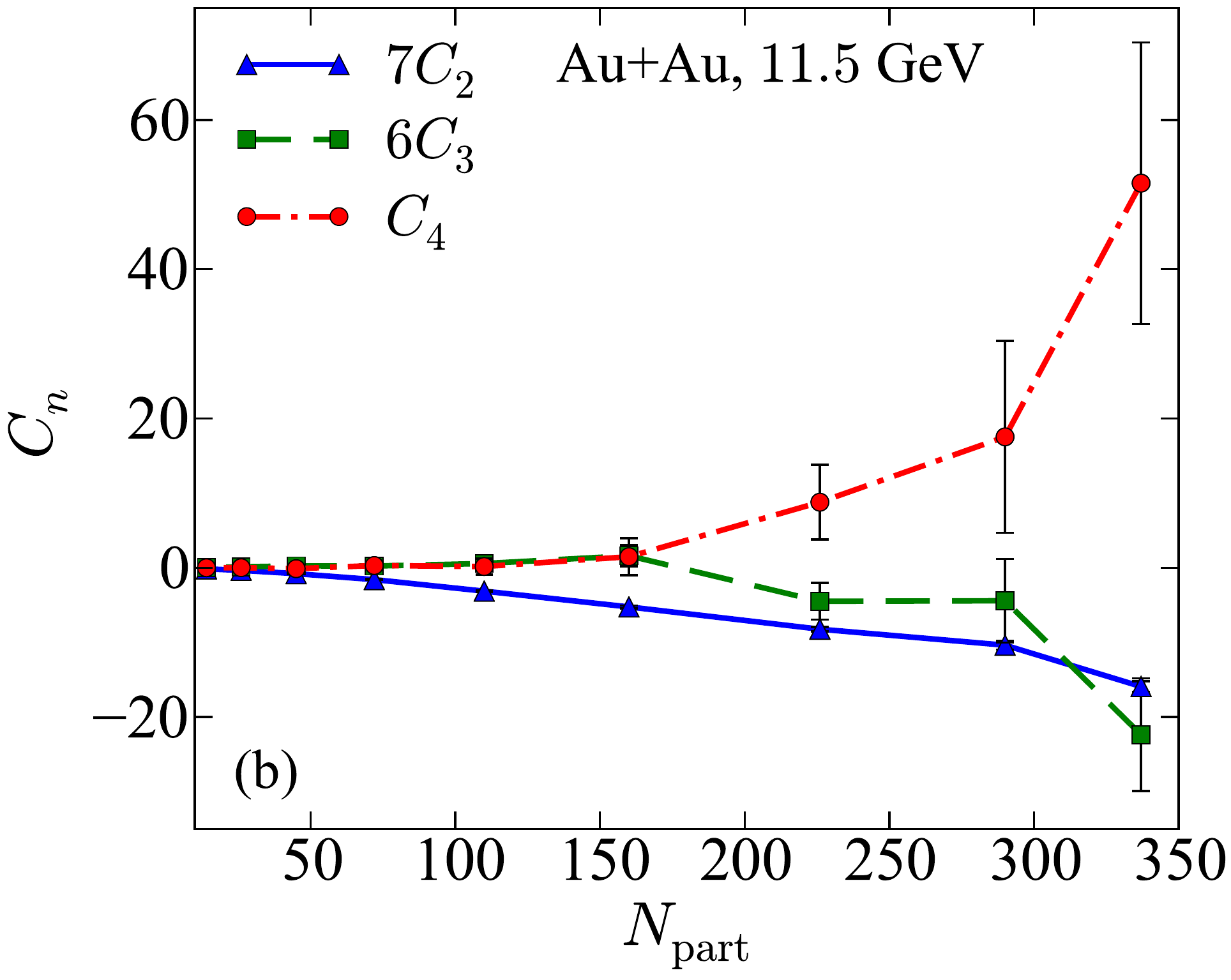}
\includegraphics[scale=0.3]{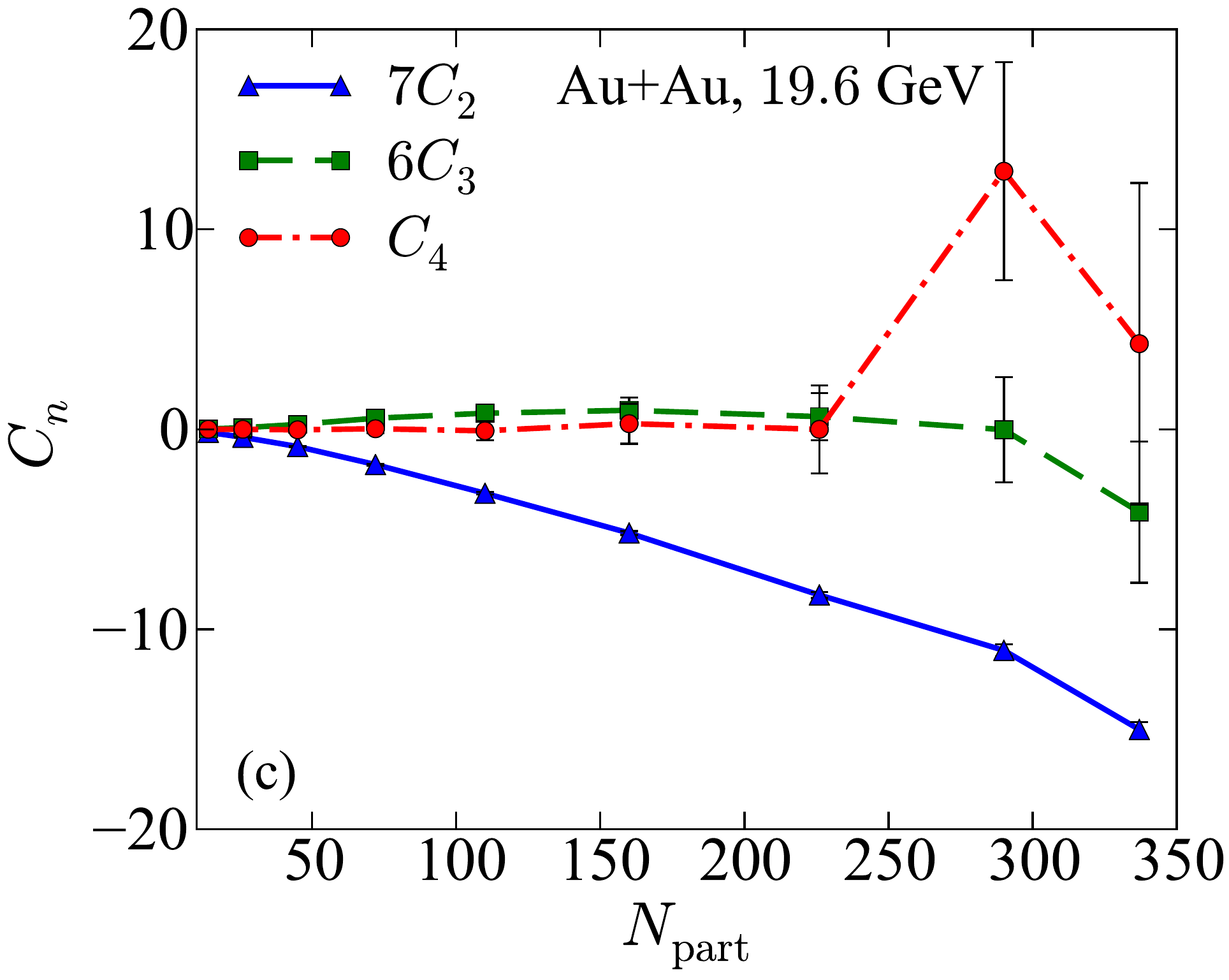}
\end{center}
\par
\vspace{-5mm}
\caption{Centrality dependence of the two- three- and four-proton
  correlation functions $C_{2},C_{3},C_{4}$ for collision energies
  $\sqrt{s}=7.7\gev$ (a), $11.5\gev$ (b), and
  $19.6\gev$ (c). Results are based on preliminary STAR data \cite{Luo:2015ewa}.}
\label{fig:capital}
\end{figure} 

% Full correlations
Having defined the correlation functions and their relation to the cumulants
we can now proceed to extract them from the measured proton number cumulants
obtained by the STAR collaboration \cite{Adamczyk:2013dal,Xu:2014jsa,Luo:2015ewa}. Here we will concentrate on the preliminary data which cover the transverse momentum range $0.4\gev<p_{t}<2.0\gev$ \cite{Luo:2015ewa}.  

Our goal is to extract information about correlations between
  protons, given by the genuine multi-proton integrated correlation
  functions, $C_n$ and $c_n$, using the measured cumulants for protons
  (not net-proton). It would be also interesting to extract analogous
  information about the antiproton correlation functions (or even
  proton-antiproton correlations, see Appendix), however, in this
  paper we are interested in the lowest beam energies, where the
  number of antiprotons is small.\footnote{For example, in the
    most central Au+Au collisions at $7.7$ GeV, the average number of
    measured antiprotons in $|y|<0.5$ approximately equals $0.3$
    compared to roughly $40$ protons.}
Let us add here that the cumulant ratio, $K_4/K_2$, is very similar for protons and net-protons for all measured energies, see a recent review \cite{Luo:2017faz}, including $7.7$ GeV where $K_4/K_2$ is particularly large.\footnote{At low energies this is rather obvious since contribution from antiprotons to cumulants is suppressed by powers of the number of antiprotons, see Eq. (A6) in the Appendix.}

Let us start with the correlation functions $C_{n}$,
Eqs.~(\ref{eq:C2}-\ref{eq:C4}). They are shown in
Fig.~\ref{fig:capital} as a function of centrality for the three
energies, $\sqrt{s}=7.7\gev, \, 11.5\gev\, {\rm and} \,
19.6\gev$. Note that we have multiplied the correlation functions with the appropriate factors so that they reflect their contribution to the fourth order cumulant, Eq.~\eqref{eq:K4C}.
For the two most central points, we find that for all three energies
the four-proton correlations are finite and positive,
 $C_{4}>0$, whereas the two- and three-proton correlations are negative, $C_{2},C_{3}<0$. In addition, for $\sqrt{s}=7.7\gev$ $C_{4}$ is clearly the dominant contribution to the fourth order cumulant. Thus, the steep rise in the $K_{4}/K_{2}$ cumulant ratio seen in the preliminary STAR data \cite{Luo:2015ewa} is indeed due to four-proton correlations. For $\sqrt{s}=19.6\gev$ on the other hand we find that for the most central point the {\em negative} two-particle correlation is the dominant contribution to the fourth order cumulant. Therefore, the fact that the preliminary STAR data show a cumulant ratio below the Poisson baseline, $K_{4}/K_{2}<1$, is due to negative two-proton  rather than negative four-proton correlations.

\begin{figure}[h]
\begin{center}
\includegraphics[scale=0.29]{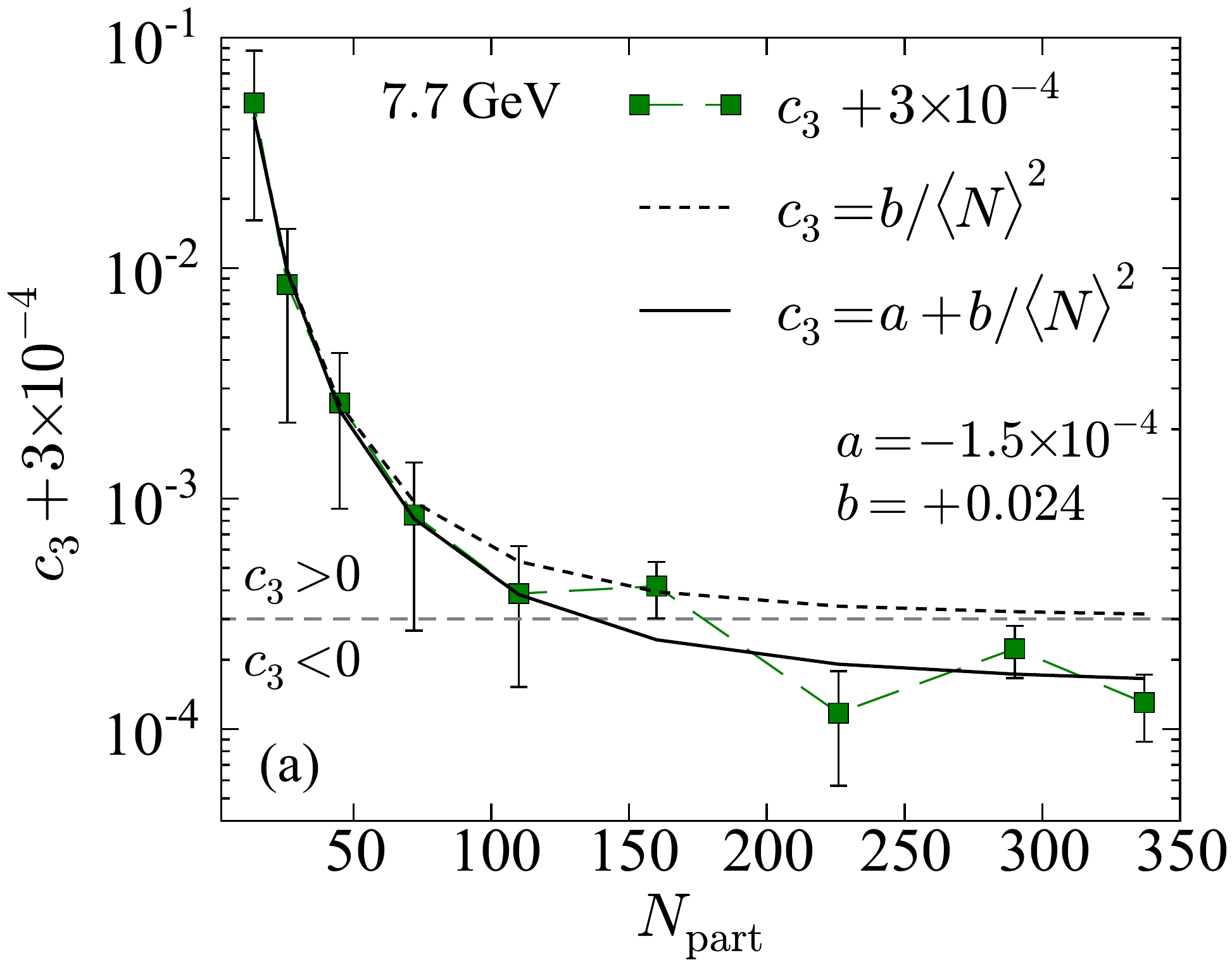}
\includegraphics[scale=0.29]{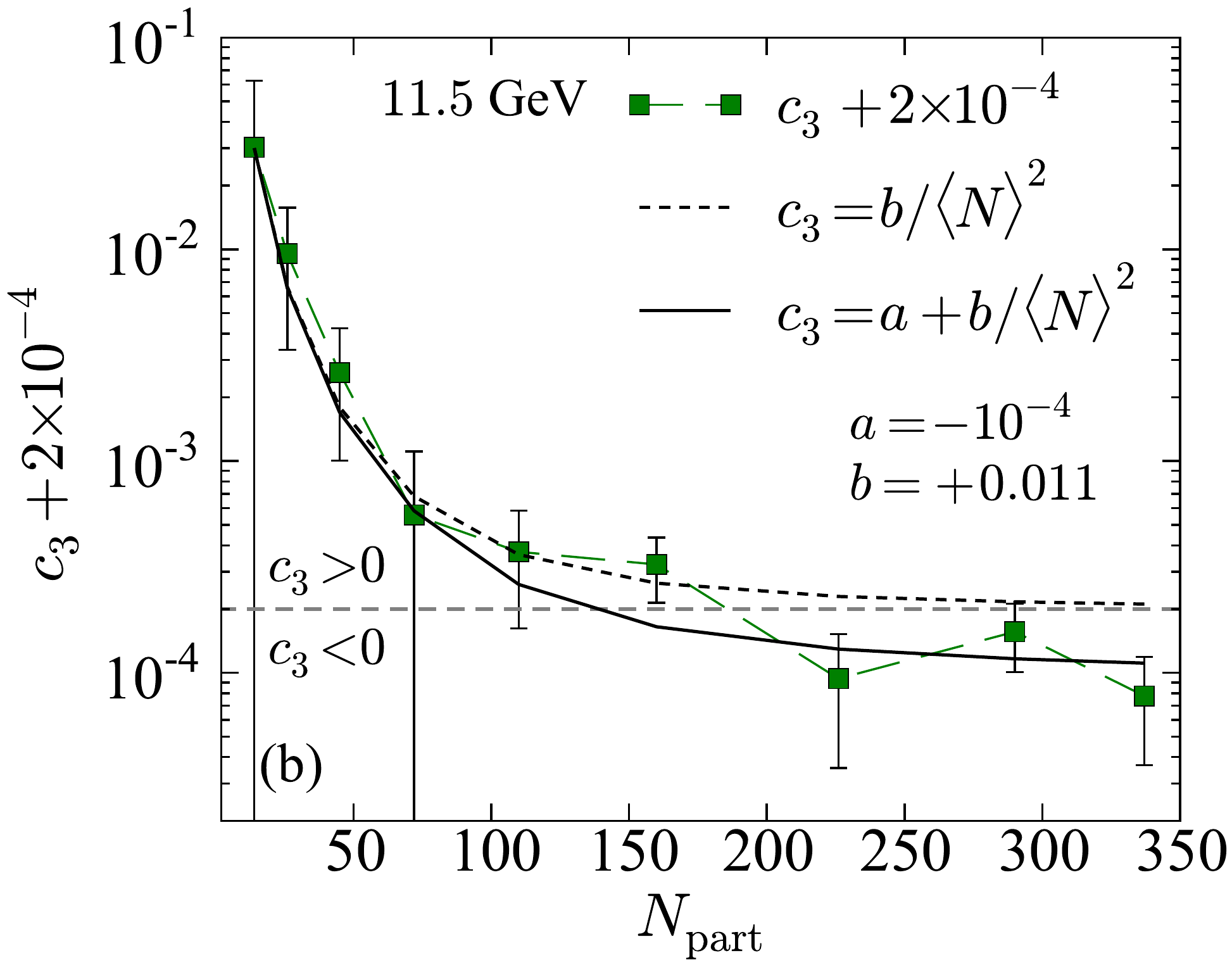}
\includegraphics[scale=0.29]{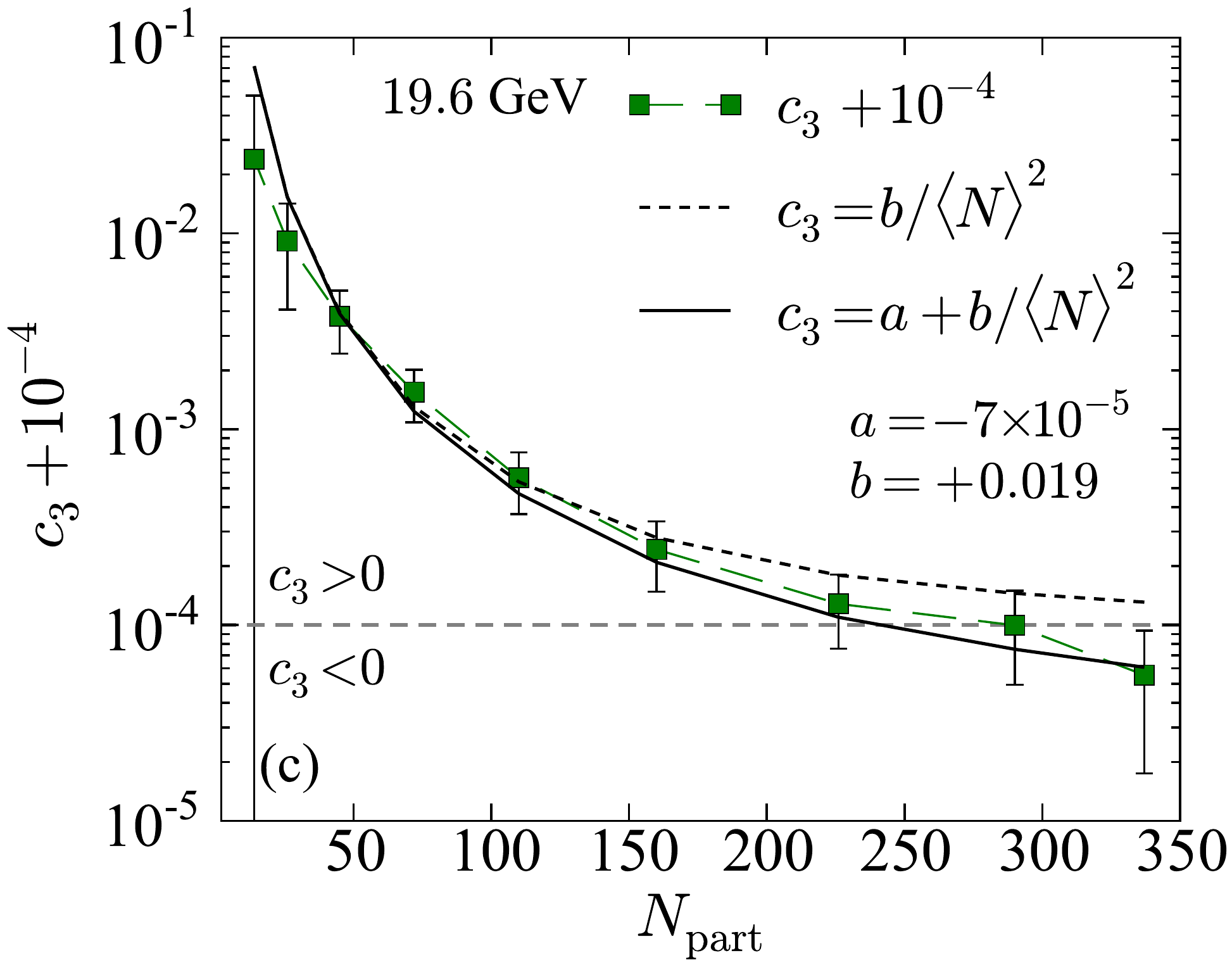}
\end{center}
\par
\vspace{-5mm}
\caption{Centrality dependence of three-proton reduced
  correlation functions $c_{3}$ for collision energies
  $\sqrt{s}=7.7\gev$ (a), $11.5\gev$ (b), and
  $19.6\gev$ (c). The horizontal long-dashed line separates
  positive from negative values. The short dashed lines represent an 
  expectation from the independent source model, $c_{3}\sim
  1/\ave{N}^{2}$ with $\ave{N}$ being the number of measured protons. 
  The full lines adds a constant
  offset to the independent source model.
  Results are based on preliminary STAR data \cite{Luo:2015ewa}.}
\label{fig:c3}
\end{figure} 

\begin{figure}[t]
\begin{center}
\includegraphics[scale=0.29]{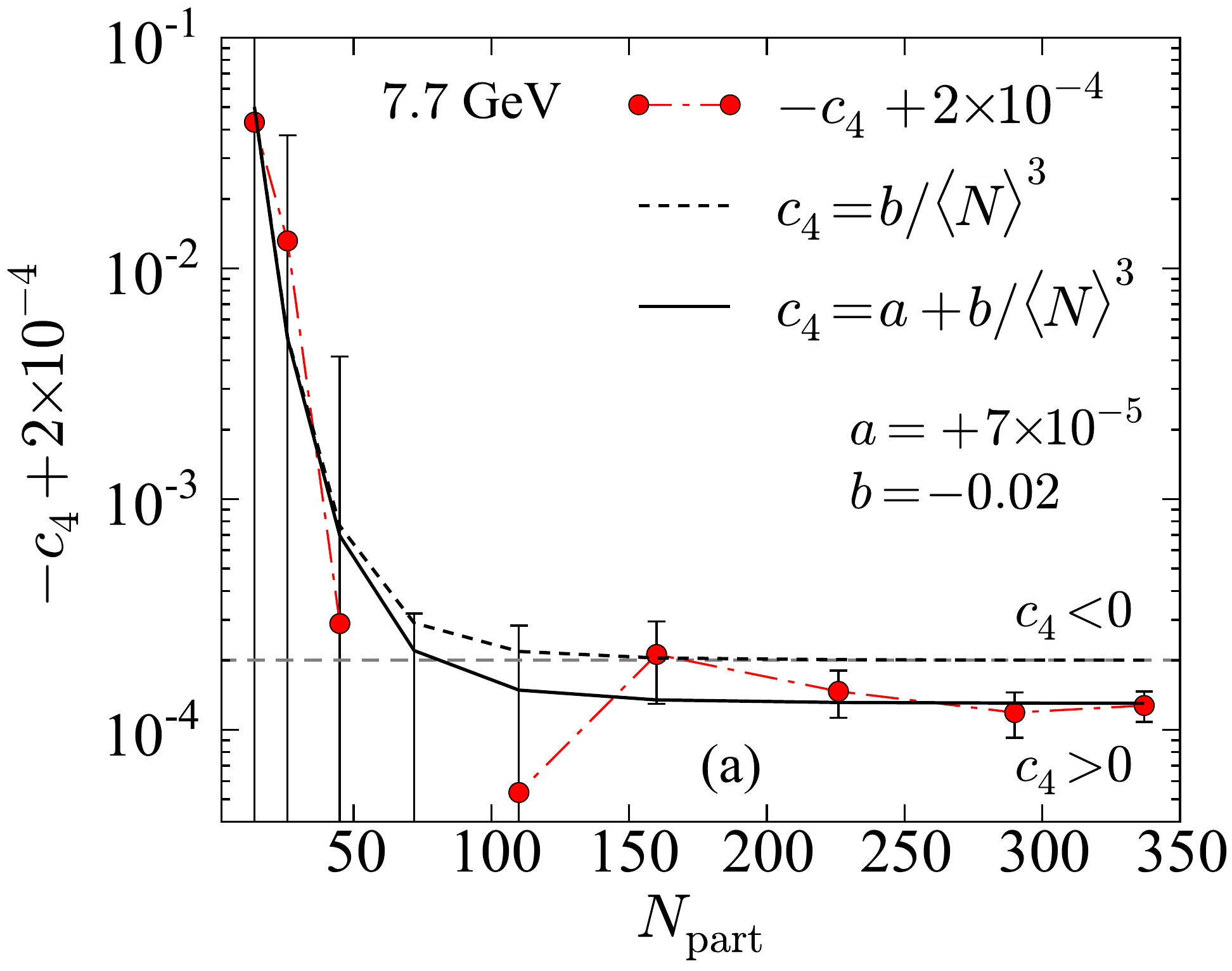}
\includegraphics[scale=0.29]{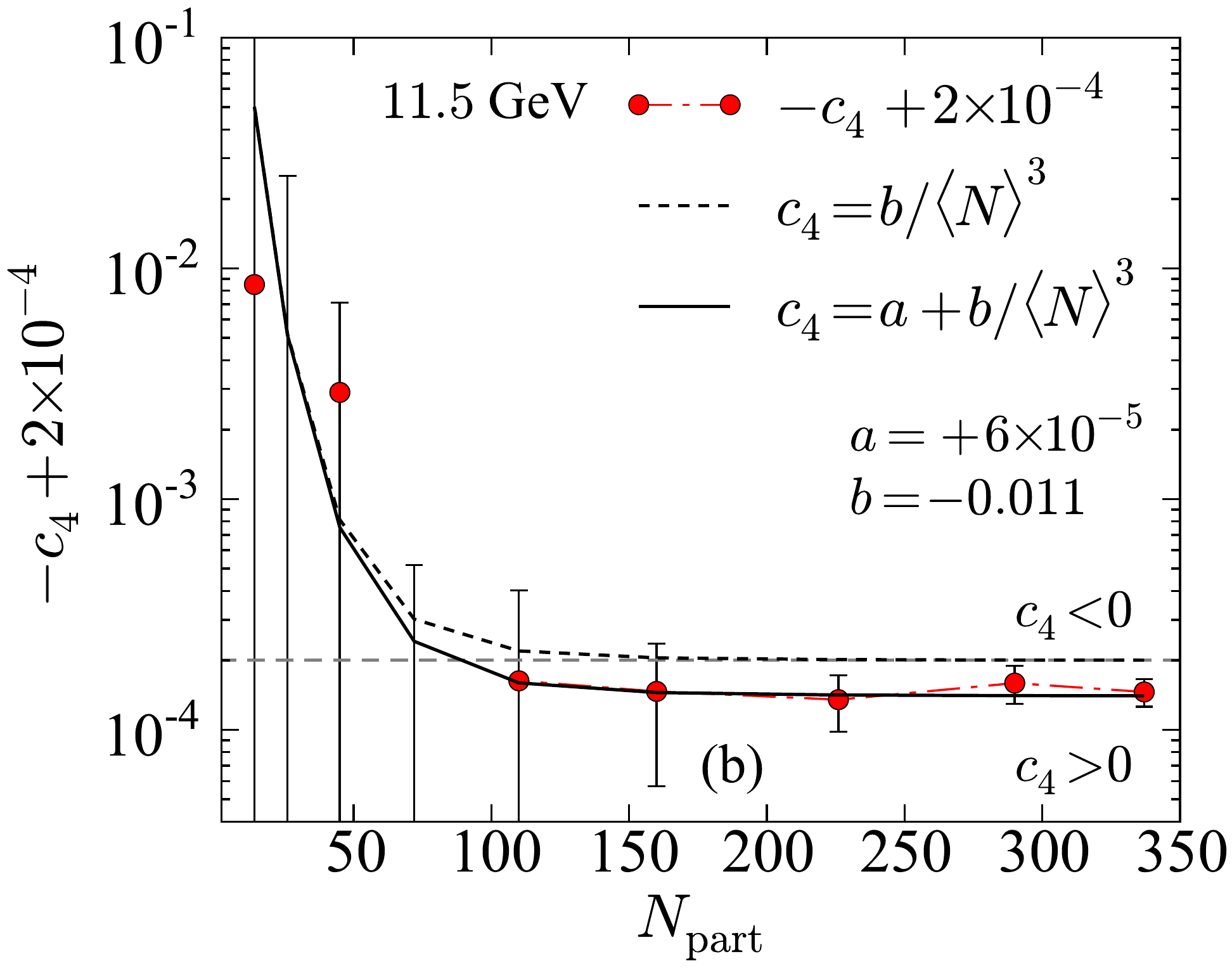}
\includegraphics[scale=0.29]{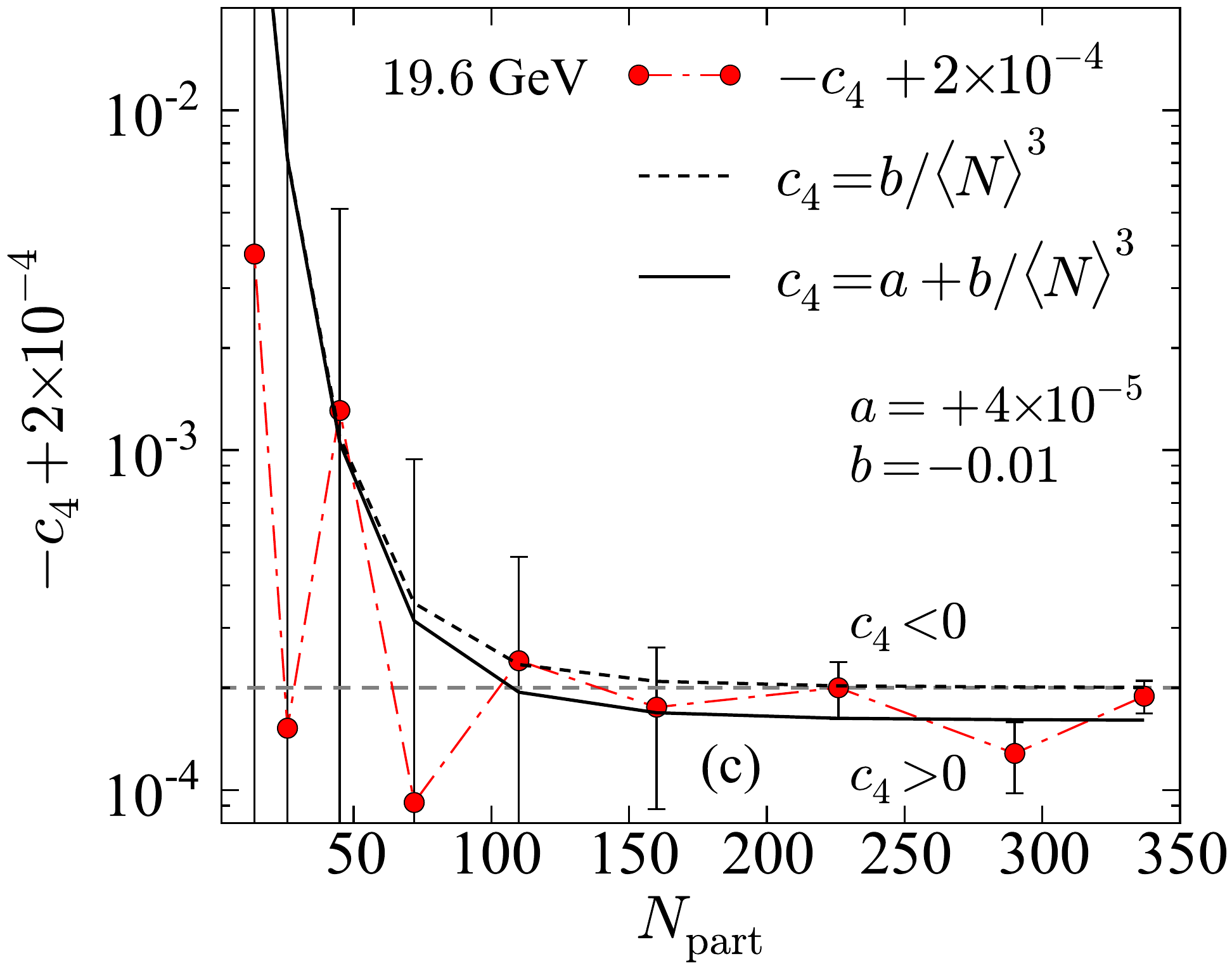}
\end{center}
\par
\vspace{-5mm}
\caption{Same as Fig.~\ref{fig:c3} but for the four-proton reduced
         correlation function $c_{4}$.} 
\label{fig:c4}
\end{figure}

\begin{figure}[b]
\begin{center}
\includegraphics[scale=0.3]{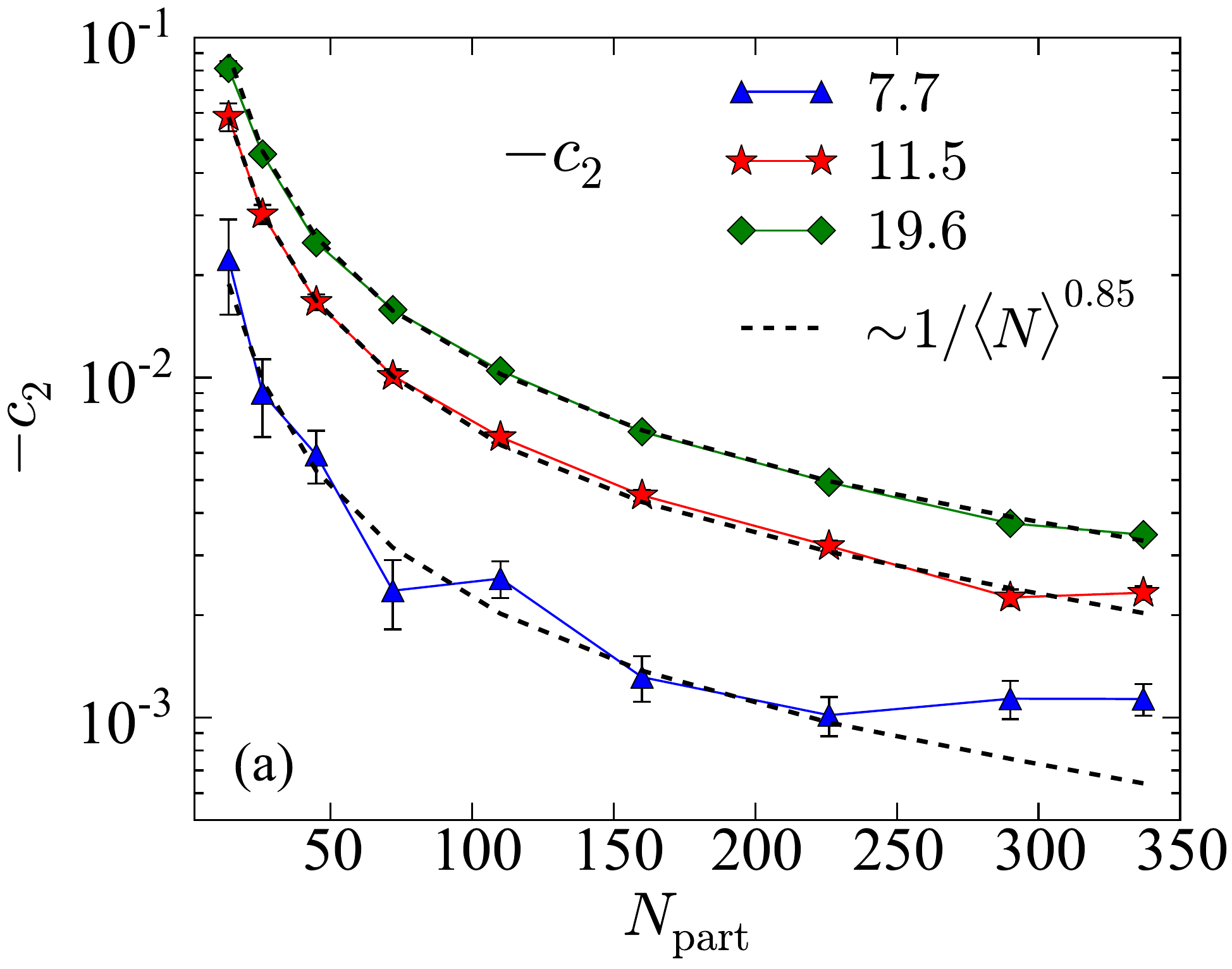}
\includegraphics[scale=0.3]{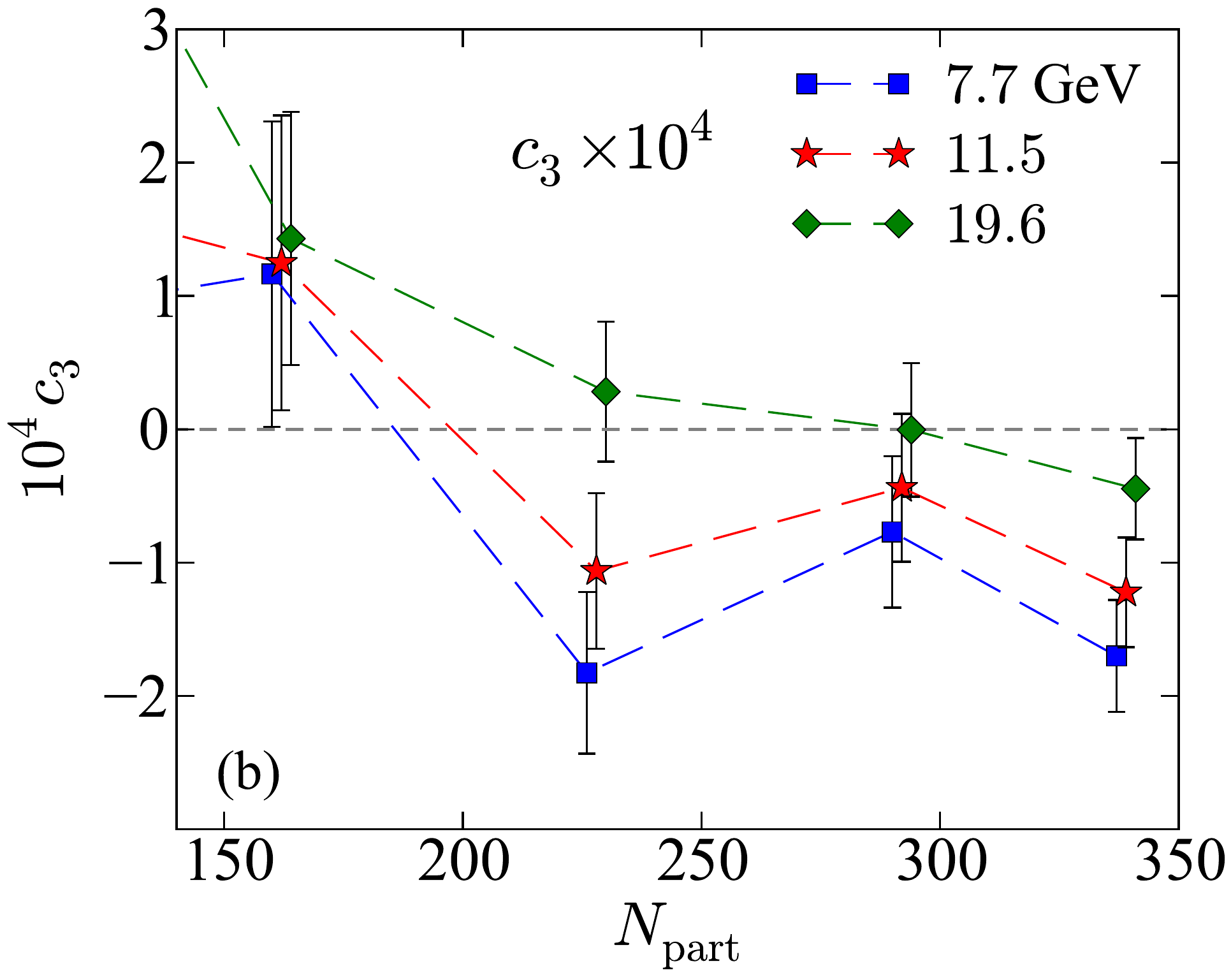}
\includegraphics[scale=0.3]{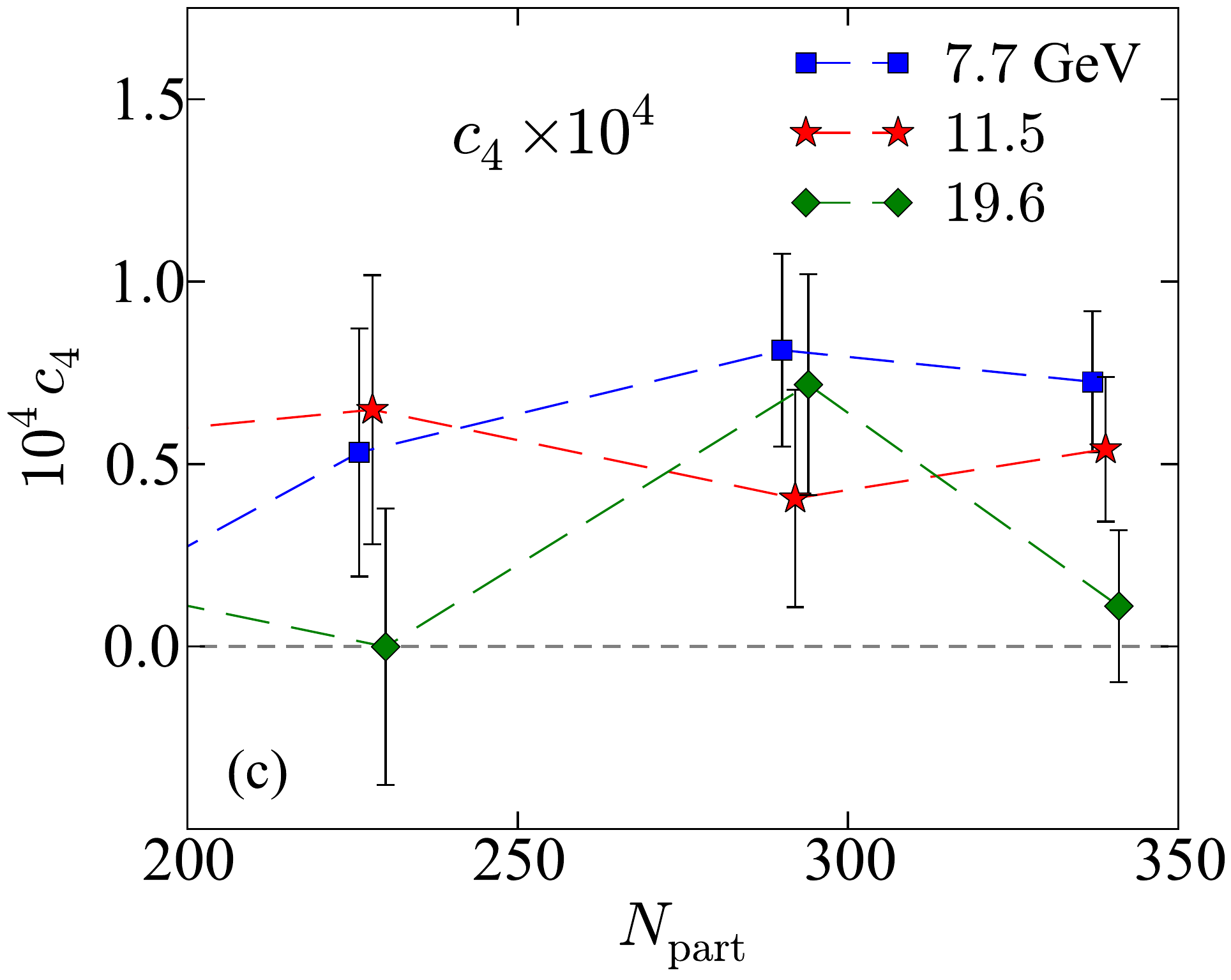}
\end{center}
\par
\vspace{-5mm}
\caption{Panel (a): Centrality dependence of two-proton reduced
  correlation functions $c_{2}$ for collision energies
  $\sqrt{s}=7.7\gev$, $11.5\gev$, and $19.6\gev$. The dashed lines are to guide the eye and demonstrate that  
  $c_{2}\sim 1/\ave{N}^{0.85}$, where $\ave{N}$ is the number of protons. Panel (b): Most
  central points for the three-proton reduced correlation $c_{3}$ function for
  energies $\sqrt{s}=7.7\gev$, $11.5\gev$, and $19.6\gev$. Panel (c): 
  Most central points for the four-proton reduced correlation
  function $c_{4}$ for
  energies $\sqrt{s}=7.7\gev$, $11.5\gev$, and $19.6\gev$. 
  Results are based on preliminary STAR data \cite{Luo:2015ewa}.}
\label{fig:energy}
\end{figure}

%reduced correlations
Next we turn to the  reduced correlation functions, $c_{2},c_{3},c_{4}$, Eq.~\eqref{eq:coupling}. In Figs. \ref{fig:c3}, \ref{fig:c4} and ~\ref{fig:energy} (panel (a)) we show their centrality dependence for the three energies under consideration. We find that the reduced two-proton correlations or couplings, $c_{2}$, for all energies scale like $1/\ave{N}^{0.85}$, with $\ave{N}$ being the number of protons, which is close to the $1/\ave{N}$ scaling expected from independent sources, but sufficiently different that this behavior deserves further investigation. At present we have no obvious explanation for this deviation from independent source scaling.

For $N_{\rm part}<200$ the three- and four-proton couplings, within errors, are consistent with $1/\ave{N}^{2}$ and $1/\ave{N}^{3}$ scaling, respectively. In addition the three- and four-proton couplings change sign around 
$N_{\rm part}\simeq 200$ whereas $c_{2}$ remains negative for all centralities. At roughly the same centrality, the three- and four-proton couplings flatten out, most prominently at the lowest two energies. 
Concentrating on the lowest energy, $\sqrt{s}=7.7\gev$, we find that for $N_{\rm part}>200 $ all three reduced correlation functions remain constant, indicating stronger correlations than an independent source picture would suggest. As discussed above, this ``collective'' behavior  may be due to either correlations among the sources or due to sources which correlate increasingly more particles (e.g., clusters increase their particle content with increasing $N$). It is interesting to note, that the transition from independent source scaling to ``collective'' behavior is accompanied with a change of sign of the three- and four-proton couplings. Apparently some new dynamics comes into play at $N_{\rm part}\simeq 200$. The two right panels of Fig.~\ref{fig:energy} show this region in more detail. It appears that the centrality independence is most significant for the lower energies, whereas it would be difficult to argue for a centrality independence, especially for $c_{3}$, at $19.6\gev$. 

We note that the purpose of the solid and dashed lines presented in Figs. \ref{fig:c3}, \ref{fig:c4} and ~\ref{fig:energy} (panel (a)) is to guide the eye and demonstrate that the preliminary STAR data are roughly consistent (except most central collisions) with $c_k \sim 1/\left\langle N\right\rangle^{k-1}$ expected from the independent source model, where $\left\langle N\right\rangle$ is the number of protons at a given centrality. Finally, let us also add that according to the preliminary STAR data $\ave{N} \sim N_{\rm part}^{1.25}$, which allows to translate the number of protons to the number of participants.

\begin{figure}[t]
\begin{center}
\includegraphics[scale=0.3]{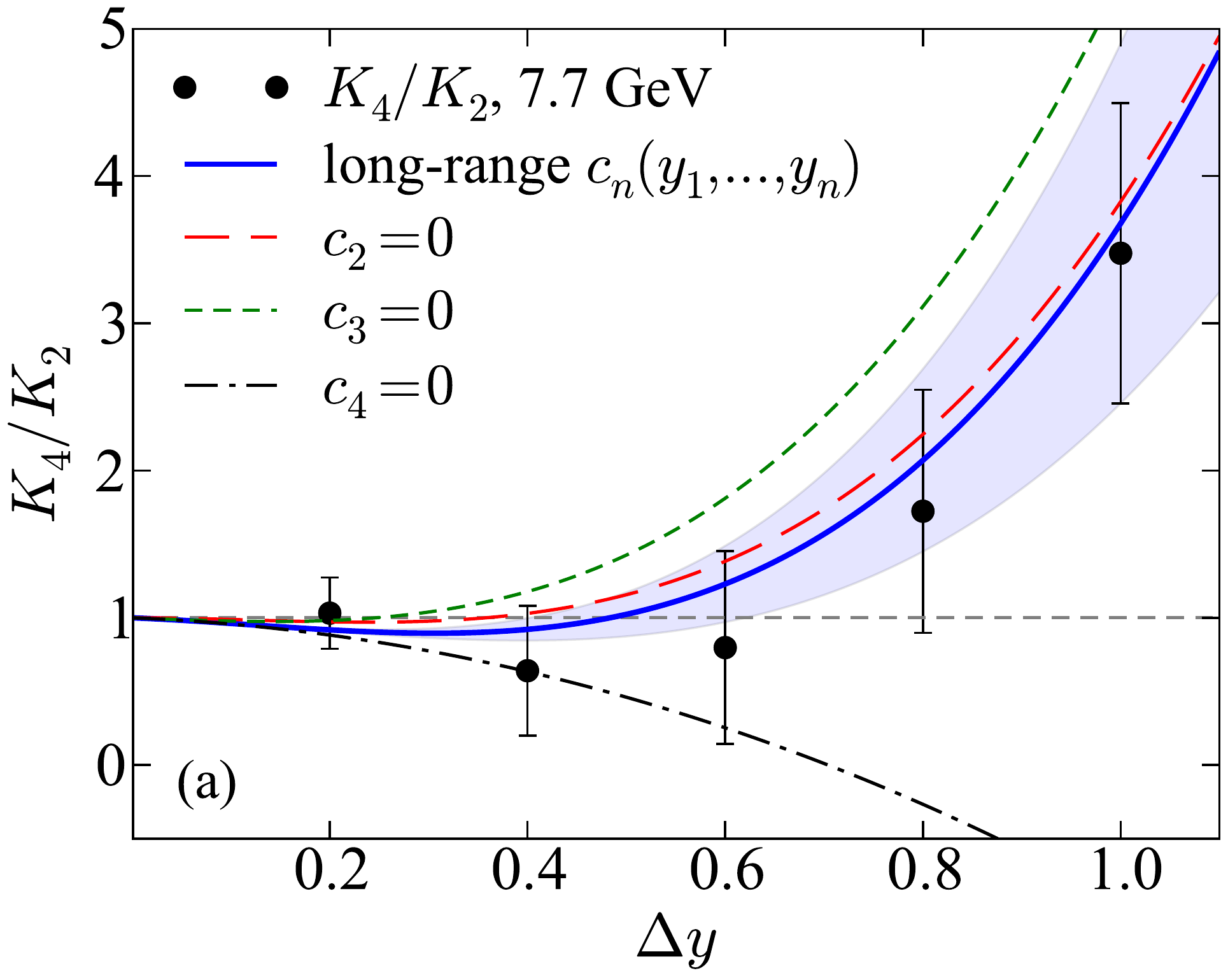} 
\hspace{1cm}
\includegraphics[scale=0.3]{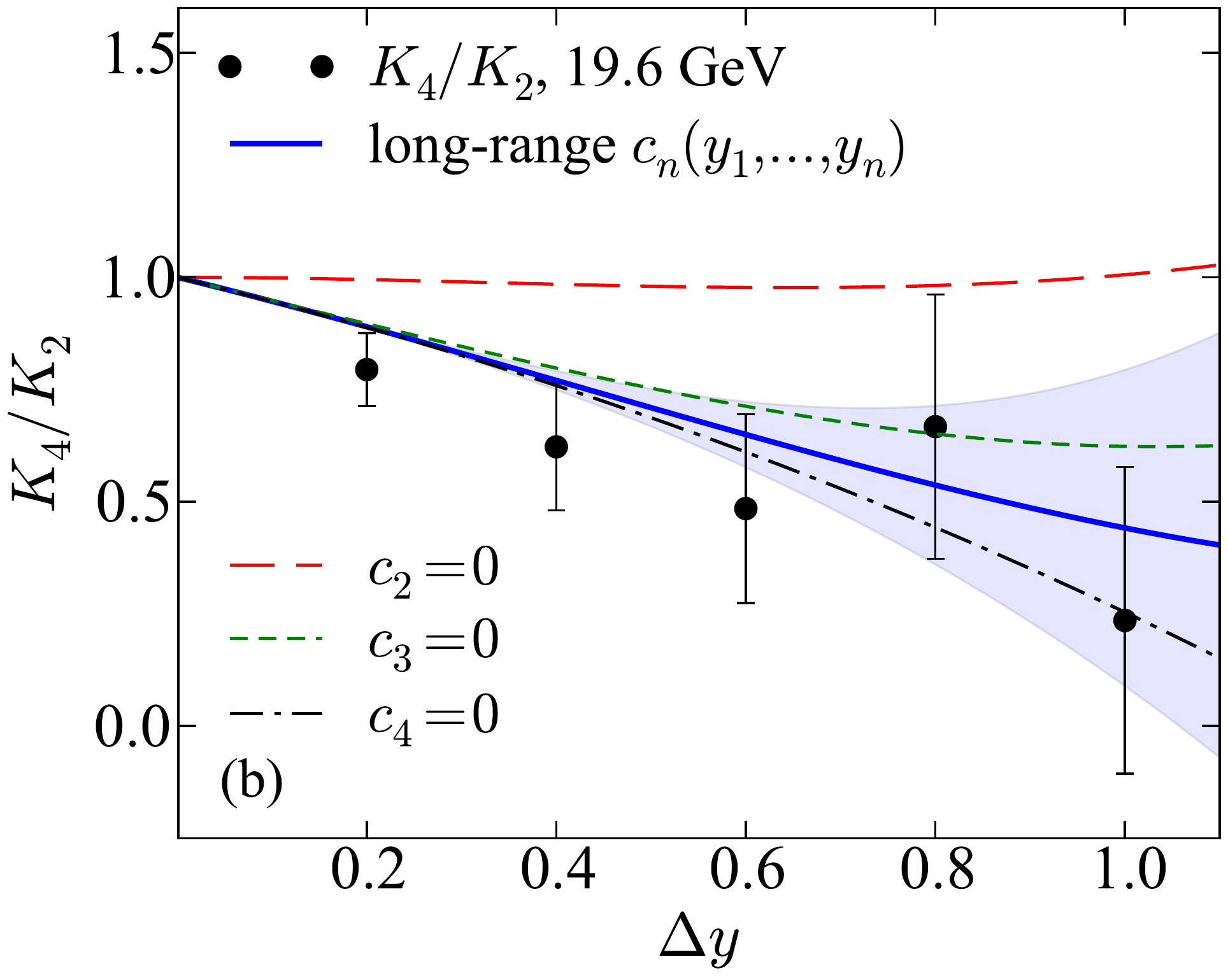}
\end{center}
\par
\vspace{-5mm}
\caption{Dependence on the rapidity window $\Delta y$ of the cumulant
  ratio $K_{4}/K_{2}$. The full line corresponds to our prediction
  assuming long-range correlations (see text fore details). The shaded area represent the error on this
  prediction. The long-dashed (red), short-dashed (green) and
  dot-dashed (black) curve correspond to setting $c_{2}=0$ or
  $c_{3}=0$ or $c_{4}=0$, respectively. Panel (a) is for
  $\sqrt{s}=7.7\gev$ and panel (b) for $\sqrt{s}=19.6\gev$.
  The data are preliminary STAR results \cite{Luo:2015ewa}.}
\label{fig:rapi}
\end{figure}

%%% rapidity dependence
Next let us turn to the  dependence of the cumulant ratio
$K_{4}/K_{2}$ on the size of the rapidity window $\Delta y$, 
where protons are accepted. Preliminary results of this ratio has been
shown by STAR for rapidity windows up to $\Delta y \le 1$
\cite{Luo:2015ewa}. As discussed in section~\ref{sec:comments} the
cumulant ratio $K_{4}/K_{2}$ is constant in case of short range
correlations in rapidity. For long range correlations, on the other
hand, the dependence of the cumulants on the rapidity is a given by a
polynomial of up to $n^{\rm{th}}$ order, where $n$ is the order of
the cumulant. In Fig.~\ref{fig:rapi} we show the preliminary STAR data
\cite{Luo:2015ewa} for both $7.7\gev$ and $19.6\gev$ together with the
results assuming long range correlations.\footnote{We remind the
  reader that short- and long-range is relative to the rapidity bin
  under consideration. At present the maximum rapidity bin is $\Delta
  y = 1$ which is rather modest.} Clearly the STAR data show a
significant dependence on $\Delta y$, ruling out short-range
correlations. The predictions based on long-range correlations, on the
other hand, agree with the preliminary STAR data rather well.\footnote{The most direct way to verify the long-range character of the observed correlations is to measure $C_{n}$ for different values of $\Delta y$ and see if they satisfy the relation given by Eq. (\ref{eq:Cn-Delta-y}). }

The blue solid lines in Fig.~\ref{fig:rapi} were generated using
  Eq. (\ref{eq:K4-rapi}) for $K_4$ and an analogous expression for $K_2$. From the preliminary STAR data we have $\langle N_{\Delta y=1}\rangle = 39.3$ for central $7.7$ GeV and $24.9$ for $19.6$ GeV. Using $\langle N\rangle = \langle N_{\Delta y=1}\rangle \Delta y$ we obtain the rapidity dependence of $\langle N\rangle$. In case of long-range correlation there are three unknown parameters $c_{n}^{0}$ for $n=2,3,4$. We determine them using the STAR values of $K_{2,3,4}$ at $\Delta y=1$, which allow to calculate $C_n$ and consequently $c_n=c_{n}^0$. Having $c_{n}^{0}$ (determined from $\Delta y=1$) we can calculate $K_4/K_2$ for arbitrary values of $\Delta y$.\footnote{We make our calculation using the preliminary STAR cumulants for protons at $\Delta y=1$, however in Fig.~\ref{fig:rapi} we compare to the rapidity dependence of net-proton data (the only data currently available on rapidity dependence). It explains a slight disagreement at $\Delta y=1$, which is obviously more visible at $19.6$ GeV. } 

In Fig.~\ref{fig:rapi} we also
show the resulting rapidity dependence when we set one of the
couplings to zero. For $7.7\gev$ setting  $c_{2}=0$ makes hardly any
difference and even $c_{3}=0$ bring the result close within
errors. Clearly, as already emphasized, the ratio $K_{4}/K_{2}$ for
central $7.7\gev$ collisions is dominated by four-proton
correlations. This is different for $19.6\gev$ shown in panel (b). The $K_{4}/K_{2}$ ratio drops more or less linearly with
$\Delta y$. This dependence suggests that the second order correlation
function  $C_{2}$ dominates the cumulant ratio, and that $C_{2}$ is
negative. This is quantified by our results. While $c_{3}=0$ or
$c_{4}=0$ still gives reasonable agreement, setting $c_{2}=0$ totally
misses the data. This observation supports our previous finding that
the drop in the cumulant ratio below the Poisson limit,
$K_{4}/K_{2}<1$ at $19.6\gev$ originates from  negative two-proton correlation.

\begin{figure}[h]
\begin{center}
\includegraphics[scale=0.35]{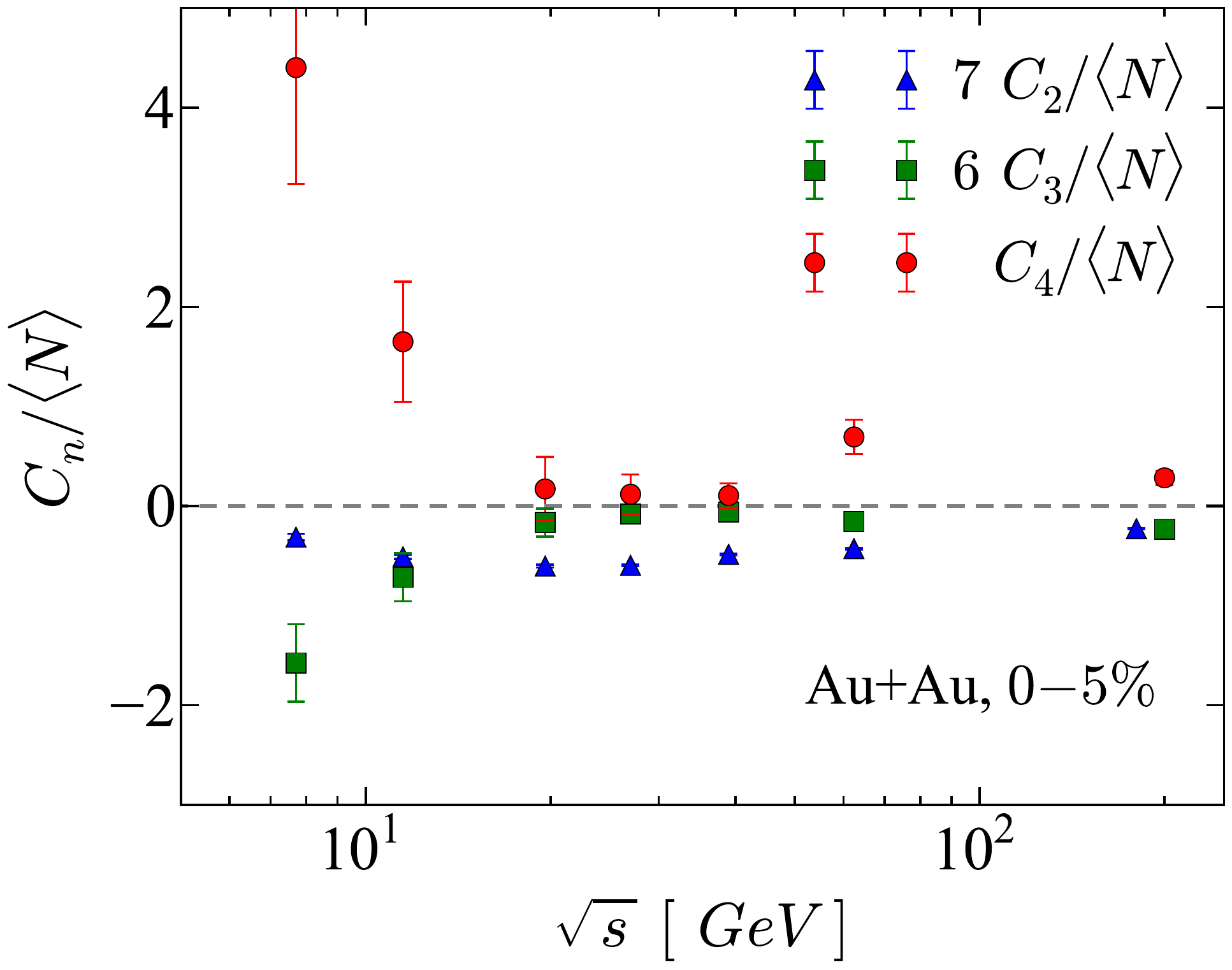} 
\end{center}
\par
\vspace{-5mm}
\caption{Energy dependence for the scaled correlation functions
  $C_{n}/\ave{N}$ weighted with the appropriate factor to reflect the
  relative contribution to the fourth order cumulant $K_{4}$ (see
  Eq.~\eqref{eq:K4C}). The 200 GeV point for $C_2$ is shifted for clarity.
  Results are based on preliminary STAR data \cite{Luo:2015ewa}.}
\label{fig:energy_dependence}
\end{figure}

Finally in Fig.~\ref{fig:energy_dependence} we show the energy
dependence of the cumulants $C_{n}$ which we scaled by the
number of particles, $C_{n}/\ave{N}$ and multiplied by the appropriate
factors to reflect their relative contribution to the fourth order
cumulant, Eq.~\eqref{eq:K4C}. Here we include points for the proton
correlations up to $\sqrt{s}=200\gev$ to show the overall trend 
although at energies larger that $19.6\gev$  anti-protons become
non-negligible, and thus the physical interpretation is less clear. 
In spite of that there seems to be a clear trend as we lower the
energy. Aside from an excursion at $62.4\gev$ the scaled four-particle
correlation seems to be small, slightly positive before it
significantly increases for the two lowest energies. Clearly the
excursion at $62.4\gev$ needs further scrutiny. Similarly, the scaled
three-proton correlation stays flat and negative before it decreases
even further at the lowest energy. The scaled two-particle correlation,
on the other hand, seem to exhibit a shallow minimum around
$20-30\gev$. At lower energies it tends towards zero, and one might be
inclined to
speculate that it may turn positive at even lower energies. Needless
to say, the 
strong energy dependence of the three- and four-proton correlations
together with the prospect of the two particle correlation changing sign
warrants measurements at even lower energies.  

To summarize this section, we have used the preliminary STAR data on
proton cumulants to extract information about the correlation
functions and couplings for protons. We find that at the lowest beam energy of
$7.7\gev$ there are significant four-proton correlations which
dominate the fourth order cumulant. At $19.9\gev$, on the other hand,
the fourth order cumulant is dominated by a negative two-proton
correlation. We further observed that for the lowest energies the
centrality dependence change from  that of an independent source model to a ``collective'' one at $N_{\rm part}\simeq 200$. At about the same centrality the three- and four-proton couplings change sign, indicating a change in the underlying dynamics. Finally, an analysis of the rapidity dependence indicates that the correlations are long-range in rapidity. Of course given the fact that $\Delta y \le 1$ we can not rule out a finite correlation length which is somewhat larger that $\delta y = 1$. The rapidity dependence also confirms  our finding that the cumulant ratio $K_{4}/K_{2}$ is dominated by (positive) four-proton correlations at $7.7\gev$ and by (negative) two-proton correlations 
at $19.6\gev$. Finally, the scaled correlations show interesting
dependence on the energy especially at the lowest available energies,
which clearly calls for measurements at even lower energies that $\sqrt{s}=7.7\gev$.

\section{Potential implications for the search of a critical point}
In this section we want to explore to which extent the signs of the
correlation functions $C_{2},C_{3},C_{4}$  
can be used to exclude regions around a QCD critical point. 
Here we use universality arguments in 
analogy to \cite{Nonaka:2004pg,Stephanov:2011pb} exploiting the fact
that the critical point belongs to the Ising universality class. 
This exercise should be  considered a feasibility
study with the aim to demonstrate that the signs of the
correlation functions provide already very important information. 
For a more quantitative analysis of experimental data, additional
effects need to be accounted for 
\cite{Hippert:2015rwa}.
For example it is known that the
non-equilibrium effects can significantly alter the cumulants and
the correlation functions
\cite{Kitazawa:2013bta,Mukherjee:2015swa,Herold:2016uvv}. These
effects and possibly others, need to be corrected for in an analysis of experimental data
before any comparison with an equilibrium phase diagram is
possible. We emphasize again that our goal here is simply
to demonstrate that the signs of the correlation functions carry
non-trivial information and, when all effects are taken into account
(if at all possible), could be used to exclude certain regions of
the QCD phase-diagram. 

In the scaling domain density and reduced temperature
in the QCD setting can be mapped to the Ising variables reduced temperature $t$ and magnetic
field $H$. The precise mapping to the conventionally used coordinates temperature $T$
and chemical potential $\mu$ is of no relevance for this argument, we only note that
$H=t=0$ maps to the critical point and that the reduced temperature axis $t$ is tangential to the phase boundary at the critical point. 
The simplest qualitative parametrizations of freeze-out lines in terms
of Ising variables is given by $H=\text{const.}$ lines, see also the
discussion in \cite{Nonaka:2004pg,Stephanov:2011pb}.
Furthermore,  
the signs of correlation functions do not really depend on the variables used,
we avoid a discussion of the precise mapping from Ising to QCD
variables and stay with those of the Ising model.  

\begin{figure*}[t]
  \centering \subfloat[Exclusion area from $C_2<0$.\hfill\textcolor{white}{.}]{
\includegraphics[width=0.32\textwidth]{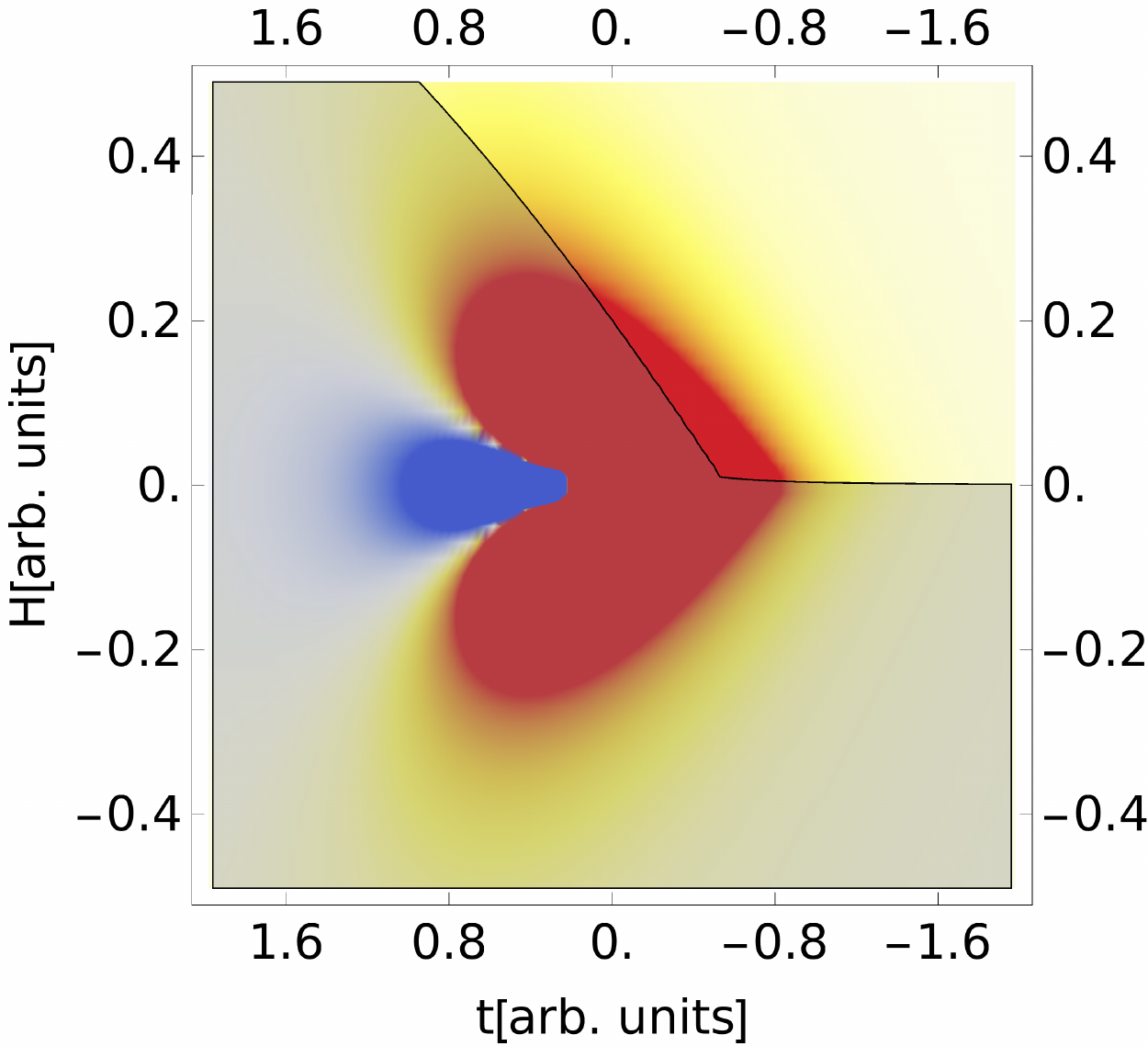}
    \label{fig:c2both}} \hfill \subfloat[Exclusion area from $C_3<0$.\hfill\textcolor{white}{.}]{\includegraphics[width=0.32
\textwidth]{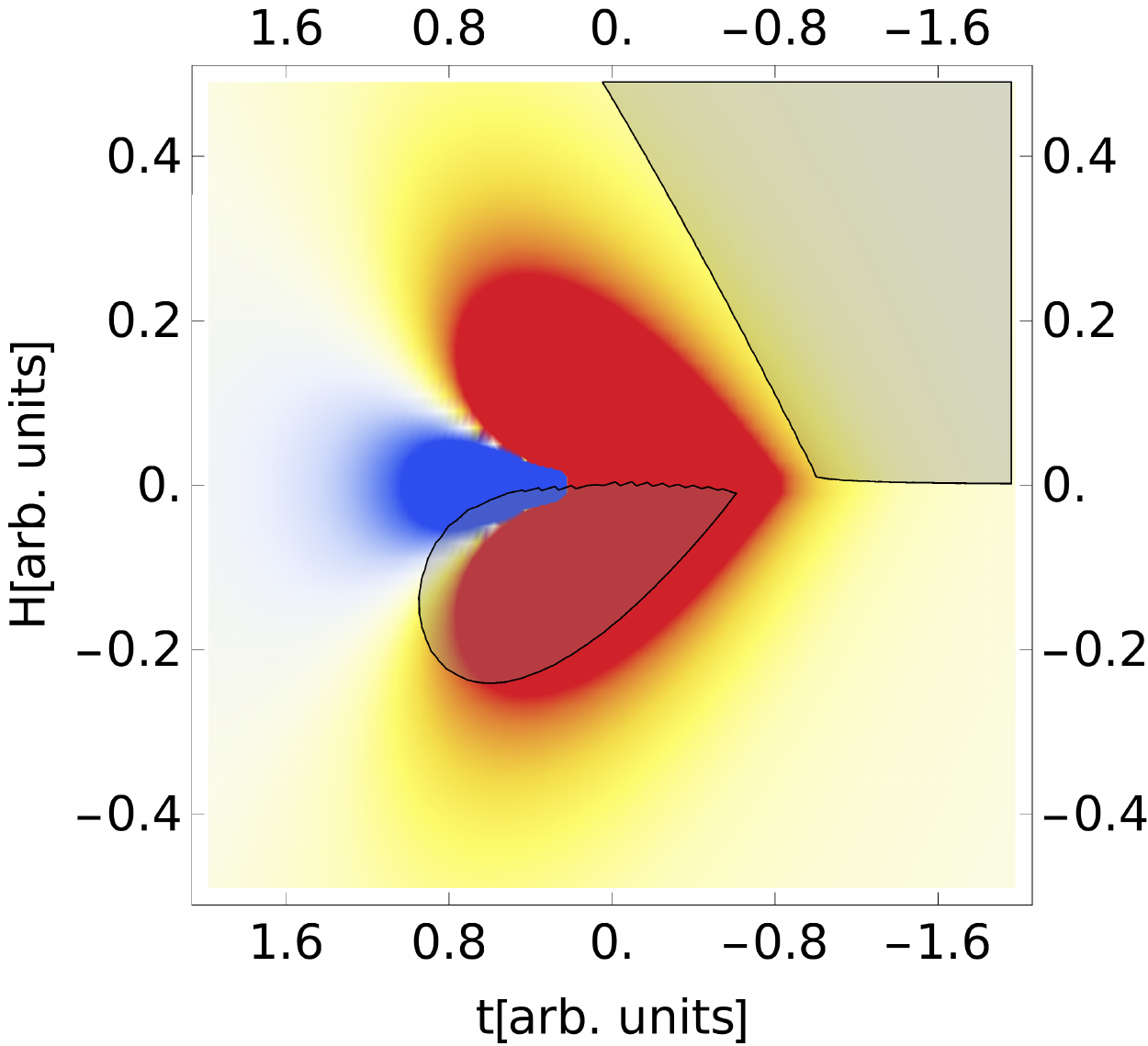}
\label{fig:c3both}}
\hfill \subfloat[Exclusion area from $C_4>0$.\hfill\textcolor{white}{.}]{\includegraphics[width=0.32
\textwidth]{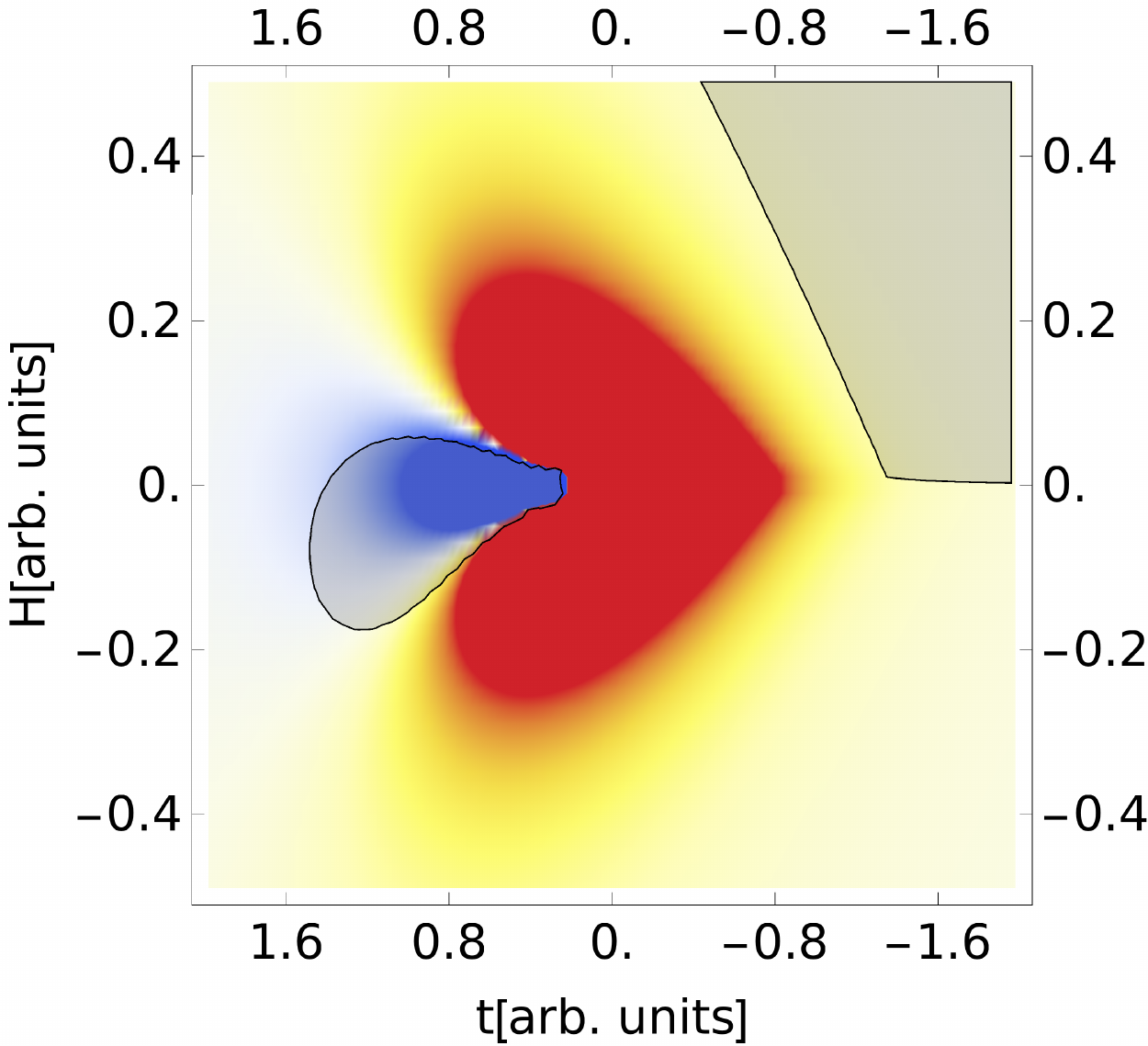}
\label{fig:c4both}}
\caption{Density plot of $K_4/K_2$ where red(blue) denotes
  positive(negative) values with excluded areas by imposing conditions
  on the signs of $C_2$, $C_3$ or $C_4$. The critical point is located
  at $H=t=0$. Excluded regions are indicated by the shaded
  areas.}
  %\hfill\textcolor{white}{.} }
\label{fig:c2c3exclusion}
\end{figure*}

We start from the standard parameterization of the magnetization $M$ in the scaling domain in
Ising variables that is given in parametric form \cite{Guida:1996ep}
\begin{equation}
\label{eq:Magnetization}
M(R,\theta)=m_0 R^\beta \theta
\end{equation}
in terms of the auxiliary variables $R$ and $\theta$ together with the relations
\begin{equation}
\label{eq:Rtheta}
t(R,\theta)=R(1-\theta^2);\quad H(R,\theta)=h_0 R^{\beta\delta}h(\theta)\,,
\end{equation}
where $m_0$ in \eq{eq:Magnetization} and $h_0$ in \eq{eq:Rtheta} denote normalization constants. 
To keep the discussion as simple as possible we employ a parameterization for $h(\theta)$ in the form of 
the linear parametric model \cite{Schofield:1969zza}, namely
\begin{equation}
h(\theta)=\theta(3-2\theta^2)\,.
\end{equation}
Now the cumulants are obtained by differentiating the magnetization
with respect to the magnetic field $H$
\begin{equation}
K_n(t,H)=\left(\frac{\partial^{n-1}M(t,H)}{(\partial H)^{n-1}}\right)_t\, ,
\end{equation}
resulting in, cf.\ \cite{Mukherjee:2015swa},
\begin{align}
K_1(t,H)&= m_0 R^{1/3}\theta\,,\nonumber\\
K_2(t,H)&= \frac{m_0}{h_0}\frac{1}{R^{4/3} (3 + 2 \theta^2)}\,,\nonumber\\
K_3(t,H)&= \frac{m_0}{h_0^2}\frac{4 \theta (9 + \theta^2)}{R^3 (-3 + \theta^2) (3 + 
   2 \theta^2)^3}\,,\nonumber\\
K_4(t,H)&= 12\frac{m_0}{h_0^3}\frac{ 81 - 783 \theta^2 + 105 \theta^4 - 5 \theta^6 +   2 \theta^8}{R^{14/3} (-3 + \theta^2)^3 (3 + 2 \theta^2)^5}\,,\label{eq:isingcumulants}
\end{align}
together with the implicit relations \eq{eq:Rtheta}. For simplicity we
inserted the approximate values $\beta=1/3$ and $\delta=5$ for the
Ising critical exponents in \eq{eq:isingcumulants}. However we
  checked that the results stay qualitatively unchanged under a
  variation of these values. Inserting \eq{eq:isingcumulants} into
  Eqs.~\eqref{eq:C2}-\eqref{eq:C4} and using $\langle N\rangle=K_1$
  (see Eq.~\eqref{eq:K1}) we
  can evaluate the couplings $C_n$ as a function of $t$ and $H$.  For
definiteness we fixed the normalization constants $m_0$ and $h_0$ by
imposing the normalization conditions $M(-1,0+)=1$ and $M(0,1)=1$. In
Fig.~\ref{fig:c2c3exclusion} we show as shaded areas which region
around the critical point is excluded by the fact that the measured
correlations functions $C_{n}$ have a certain sign. 
In addition, for orientation we also
show the regions where the cumulant ratio $K_{4}/K_{2}$ is positive
and negative (see caption for details). For suggestive reasons we
inverted the direction of the $t$-axis in all figures as in the
simplest mapping, the reduced temperature in QCD maps to the magnetic
field in Ising variables, whereas the reduced chemical potential
$\mu-\mu_c-1$ maps to the negative reduced temperature $-t$ in Ising
variables. In this way the orientation of the plots in Ising variables
can be roughly identified with the orientation of a conventional
$T-\mu$ phase diagram for QCD. In all figures, the critical point is
located at $H=t=0$. Note that whereas $K_{2N}$ ($K_{2N+1}$) is
(anti-)symmetric with respect to $H\to -H$, the couplings $C_n$ as a
sum of symmetric and antisymmetric terms no longer show this symmetry.
 
In Figs.~(\ref{fig:c2both}-\ref{fig:c4both}) we show as the shaded
areas the excluded regions around the critical point due to the
conditions $C_{2}<0$, $C_{3}<0$, and $C_{4}>0$. 
Clearly the sign of the 2-particle correlation function, $C_{2}$, imposes the
strongest constraint. In a sense this is good news, since 
the experimental determination of the two particle correlation
function requires the least statistics as it requires only the
measurement of the variance of the proton distribution. In addition,
the measurement of $C_{2}$ is less
affected by systematic uncertainties than higher order correlations, 
which probe the tails of the
distribution. Also, one would expect that the aforementioned dynamical
effects are likely to be better controlled and accounted for in case
of the two particle correlations.

In conclusion, our schematic  study showed that the signs of the correlations $C_{n}$ are
a useful tool to exclude regions in the QCD phase diagram close to the
critical point. The fact that the sign of the two particle
correlations impose the strongest constraint suggests that both
experimental as well as theoretical work should first and foremost
concentrate on the quantitative understanding of two particle
correlations.

\section{Summary and Conclusions}
In this paper we have extracted the two- three- and
four-proton correlation functions based on preliminary data
of the STAR collaboration. We have discussed how these correlation
functions are expected to scale with centrality and rapidity under
various assumptions. We found that (a) at the lowest beam energy of
$7.7\gev$ there are significant four-proton correlations. (b) At
$19.9\gev$  the fourth order cumulant is dominated by a negative
two-particle correlation. (c) For the lowest energies the centrality
dependence change from that of an independent source model to a
``collective'' one at $N_{\rm part}\simeq 200$. At roughly the same
centrality the three- and four-proton couplings change sign,
indicating a change in the underlying dynamics. (d) The preliminary
data on the rapidity dependence of the cumulant ratio $K_{4}/K_{2}$
rules out short-range rapidity correlations and is consistent with
long-range ($\Delta y >1$) correlations. (e) We looked at the energy
dependence of the relative contributions to the fourth order cumulant,
$K_{4}$ and found that, with an excursion at $62.4\gev$ the
 scaled correlation are rather constant from $200\gev$ down to
 $19.6\gev$. At lower energies both the three and four-proton
 correlations show a significant energy dependence.

We also explored to which extend the signs of the correlations
functions $C_{n}$ constraint allowed regions in the phase diagram
close tho the critical point. We found that the strongest constraint
arises from the two-particle correlations. This suggests that both
experimental as well as theoretical work should first focus on the
quantitative understanding of the two particle correlations.

Finally, we should stress that the present analysis is based on
preliminary data. Furthermore, one should not forget that there are
sources of correlations other that critical dynamics. These need to be
removed and understood, and we believe that a study of correlations
functions, preferably differential in some of their variables, 
will be essential to make progress in the search for a QCD critical point.

\begin{acknowledgments}
We thank X. Luo, V. Skokov and M. Stephanov for useful discussions and comments. 
We further thank the STAR collaboration for providing us with their
preliminary data. 
AB is supported by the Ministry of Science and Higher Education (MNiSW) and by the National Science Centre, Grant No. DEC-2014/15/B/ST2/00175, and in part by DEC-2013/09/B/ST2/00497. 
VK and NS are supported by the Director, Office of Energy Research,
Office of High Energy and Nuclear Physics, Divisions of Nuclear Physics,
of the U.S. Department of Energy under Contract No. DE-AC02-05CH11231. 
NS acknowledges funding by the DFG under grant no.\ STR 1462/1-1.
\end{acknowledgments}

\appendix

\section{Correlation functions}

Here we derive formulas for the couplings of the multi-particle genuine
correlation functions for the case of protons and anti-protons. Let $P(N,%
\bar{N})$ denotes the multiplicity distribution of protons, $N$, and
antiprotons, $\bar{N}$. The factorial moment generating function is given by%
\begin{equation}
H(z,\bar{z})=\sum\nolimits_{N}\sum\nolimits_{\bar{N}}P(N,\bar{N})z^{N}\bar{z}%
^{\bar{N}}.
\end{equation}

The factorial moments are given by%
\begin{equation}
F_{i,k}\equiv \left\langle \frac{N!}{(N-i)!}\frac{\bar{N}!}{(\bar{N}-k)!}%
\right\rangle =\left. \frac{d^{i}}{dz^{i}}\frac{d^{k}}{d\bar{z}^{k}}H(z,\bar{%
z})\right| _{z=1,\bar{z}=1} .
\end{equation}

The correlation function generating function is given by%
\begin{equation}
G(z,\bar{z})=\ln \left[ H(z,\bar{z})\right] ,
\end{equation}%
and%
\begin{eqnarray}
C_{n+m}^{(n,m)} &=&\int C_{n+m}^{(n,m)}(y_{1},\ldots ,y_{n},\bar{y}_{1},...,%
\bar{y}_{m})dy_{1}\cdots dy_{n}d\bar{y}_{1}\cdots d\bar{y}_{m}  \notag \\
&=&\left. \frac{d^{n}}{dz^{n}}\frac{d^{m}}{d\bar{z}^{m}}G(z,\bar{z})\right|
_{z=1,\bar{z}=1} ,
\end{eqnarray}%
where $C_{n+m}^{(n,m)}$ is $n+m$ correlation function with $n$ protons and $%
m $ anti-protons. When we have only protons we have $C_{n}\equiv C_{n+0}^{(n,0)}$.

Performing straightforward calculations we obtain:
\begin{eqnarray}
C_{2}^{(2,0)} &=&-F_{1,0}^{2}+F_{2,0} \\
C_{2}^{(1,1)} &=&-F_{0,1}F_{1,0}+F_{1,1}  \notag \\
C_{3}^{(3,0)} &=&2F_{1,0}^{3}-3F_{1,0}F_{2,0}+F_{3,0}  \notag \\
C_{3}^{(2,1)} &=&2F_{0,1}F_{1,0}^{2}-2F_{1,0}F_{1,1}-F_{0,1}F_{2,0}+F_{2,1} 
\notag \\
C_{4}^{(4,0)}
&=&-6F_{1,0}^{4}+12F_{1,0}^{2}F_{2,0}-3F_{2,0}^{2}-4F_{1,0}F_{3,0}+F_{4,0} 
\notag \\
C_{4}^{(3,1)}
&=&-6F_{0,1}F_{1,0}^{3}+6F_{1,0}^{2}F_{1,1}+6F_{0,1}F_{1,0}F_{2,0}-3F_{1,1}F_{2,0}-3F_{1,0}F_{2,1}-F_{0,1}F_{3,0}+F_{3,1}
\notag \\
C_{4}^{(2,2)}
&=&(-6F_{0,1}^{2}+2F_{0,2})F_{1,0}^{2}+8F_{0,1}F_{1,0}F_{1,1}-2F_{1,1}^{2}-2F_{1,0}F_{1,2}+(2F_{0,1}^{2}-F_{0,2})F_{2,0}-2F_{0,1}F_{2,1}+F_{2,2}
\notag
\end{eqnarray}
where $F_{1,0}=\langle N\rangle$, $F_{0,1}=\langle \bar{N}\rangle$. The remaining correlations $C_{n+m}^{(n,m)}$ for $m>n$ can be easily obtained by a simple change of indexes $F_{i,k} \rightarrow F_{k,i}$.

The above equations allow to express factorial moments through correlation
functions. Using formulas for the cumulants, Ref. \cite{Bzdak:2012ab}, we obtain%
\begin{eqnarray}
K_{2} &=&\left\langle N\right\rangle +\left\langle \bar{N}\right\rangle
+C_{2}^{(2,0)}+C_{2}^{(0,2)}-2C_{2}^{(1,1)} \\
K_{3} &=&\left\langle N\right\rangle -\left\langle \bar{N}\right\rangle
+3C_{2}^{(2,0)}-3C_{2}^{(0,2)}+C_{3}^{(3,0)}-C_{3}^{(0,3)}-3C_{3}^{(2,1)}+3C_{3}^{(1,2)}
\notag \\
K_{4} &=&\left\langle N\right\rangle +\left\langle \bar{N}\right\rangle
+7C_{2}^{(2,0)}+7C_{2}^{(0,2)}-2C_{2}^{(1,1)}+6C_{3}^{(3,0)}+6C_{3}^{(0,3)}-6C_{3}^{(2,1)}-6C_{3}^{(1,2)}+
\notag \\
&&C_{4}^{(4,0)}+C_{4}^{(0,4)}-4C_{4}^{(3,1)}-4C_{4}^{(1,3)}+6C_{4}^{(2,2)} 
\notag
\end{eqnarray}

Finally the reduced correlation functions or couplings are related to $C_{n+m}^{(n,m)}$ through%
\begin{equation}
c_{n+m}^{(n,m)}=\frac{C_{n+m}^{(n,m)}}{\left\langle N\right\rangle
^{n}\left\langle \bar{N}\right\rangle ^{m}} ,
\end{equation}%
or%
\begin{equation}
c_{n+m}^{(n,m)}=\frac{\int \rho _{1}\left( y_{1}\right) \cdots \rho
_{1}\left( y_{n}\right) \rho _{1}\left( \bar{y}_{1}\right) \cdots \rho
_{1}\left( \bar{y}_{m}\right) c_{n+m}^{(n,m)}(y_{1},\ldots ,y_{n},\bar{y}%
_{1},...,\bar{y}_{m})dy_{1}\cdots dy_{n}d\bar{y}_{1}\cdots d\bar{y}_{m}}{%
\int \rho _{1}\left( y_{1}\right) \cdots \rho _{1}\left( y_{n}\right) \rho
_{1}\left( \bar{y}_{1}\right) \cdots \rho _{1}\left( \bar{y}_{m}\right)
dy_{1}\cdots dy_{n}d\bar{y}_{1}\cdots d\bar{y}_{m}} .
\end{equation}

\bigskip

%\bibliography{/Users/vkoch/Documents/Bibliography/myBibliography,/Users/vkoch/Documents/Bibliography/myPapers}

%merlin.mbs apsrev4-1.bst 2010-07-25 4.21a (PWD, AO, DPC) hacked
%Control: key (0)
%Control: author (8) initials jnrlst
%Control: editor formatted (1) identically to author
%Control: production of article title (-1) disabled
%Control: page (0) single
%Control: year (1) truncated
%Control: production of eprint (0) enabled
%

\end{document}